\documentclass[prx,aps,superscriptaddress,nofootinbib,twocolumn,longbibliography,notitlepage,floatfix,10pt]{revtex4-2}

\usepackage{graphicx}
\usepackage{braket}
\usepackage{booktabs}  
\usepackage{comment}
\usepackage[normalem]{ulem}
\usepackage[title]{appendix}
\usepackage{amsmath,mathtools,amsthm,amssymb,pifont}
\usepackage[utf8]{inputenc}
\usepackage[american]{babel}
\usepackage{graphicx,xcolor,bbold,titlesec}
\usepackage{braket}
\usepackage{MnSymbol}

\usepackage[breaklinks, pdftex, hyperfootnotes=true, pdfpagelabels, bookmarks, pageanchor]{hyperref}
\hypersetup{%
	colorlinks=true, linktocpage=true, pdfstartpage=1, pdfstartview=FitH, pdfborder={0 0 0},%
	breaklinks=true, pdfpagemode=UseNone, pageanchor=true, pdfpagemode=UseOutlines,%
	plainpages=false, bookmarksnumbered, bookmarksopen=true, bookmarksopenlevel=1,%
	hypertexnames=true, pdfhighlight=/O,
	urlcolor=red, linkcolor=red, citecolor=red,
	}
\usepackage{orcidlink}
\usepackage{tikz,ifthen}
\usepackage{bbold}
\usepackage{tikz-network}
\usetikzlibrary{patterns,decorations.pathreplacing,calligraphy}
\usetikzlibrary{shapes,arrows.meta,decorations.pathmorphing}

\newcommand{\Tr}{\text{Tr}} % Trace

\newcommand{\Haar}{{\text{Haar}}} % Haar measure
\newcommand{\Gauss}{{\text{Gauss}}} % Gaussian measure
\newcommand{\Id}{\mathbb{1}}

\newcommand{\Ex}{\mathbb{E}} % Expected value
\newcommand{\Wg}{{\text{Wg}}} % Weingarten functions
\newcommand{\dist}{\mathtt{d}} % Permutation distance
\newcommand{\ad}{\mathsf{A}} % Adjacency matrix
\newcommand{\adT}{\mathsf{T}} % T matrix expansion
\newcommand{\muA}{\sigma_A} % Permutation in A

\def\scaleq{\stackrel{\rm s.l.}{=}}
\def\Sfact{S_m^{\rm fact}}

\makeatletter

\@addtoreset{paragraph}{section}  % Optional: reset numbering per section
\makeatother

\definecolor{mycolor1}{rgb}{0.82,0.82,1.}
\definecolor{mycolor2}{rgb}{0.62,0.62,1.}
\definecolor{mycolorbd}{rgb}{0.0,0.0,1.}

\begin{document}

\newcommand{\titleinfo}{
Quantum State Design and Emergent Confinement Mechanism \\ in Measured Tensor Network States
}
\title{\titleinfo}

\author{Guglielmo Lami~\orcidlink{0000-0002-1778-7263}}
\affiliation{Laboratoire de Physique Th\'eorique et Mod\'elisation, CNRS UMR 8089,
CY Cergy Paris Universit\'e, 95302 Cergy-Pontoise Cedex, France}

\author{Andrea De Luca~\orcidlink{0000-0001-7877-0329}}
\affiliation{Laboratoire de Physique Th\'eorique et Mod\'elisation, CNRS UMR 8089,
CY Cergy Paris Universit\'e, 95302 Cergy-Pontoise Cedex, France}

\author{Xhek Turkeshi~\orcidlink{0000-0003-1093-3771}}
\affiliation{Institut für Theoretische Physik, Universität zu Köln, Zülpicher Strasse 77a, 50937 Köln, Germany}

\author{Jacopo De Nardis~\orcidlink{0000-0001-7877-0329}}
\affiliation{Laboratoire de Physique Th\'eorique et Mod\'elisation, CNRS UMR 8089,
CY Cergy Paris Universit\'e, 95302 Cergy-Pontoise Cedex, France}

\begin{abstract} 
Randomness is a fundamental aspect of quantum mechanics, arising from the measurement process that collapses superpositions into definite outcomes according to Born’s rule. Generating large-scale random quantum states is crucial for quantum computing and many-body physics, yet remains a key challenge. We present a practical method based on local measurements of random Tensor Networks, focusing on random Matrix Product States (MPS) generated by two distinct quantum circuit architectures, both feasible on near-term devices.
We certify the emergent quantum randomness using the frame potential and establish a mapping between its behavior and the statistical mechanics of a domain wall particle model. In both architectures, the effect of quantum measurements induces a nontrivial confinement mechanism, where domain walls are either trapped by an external potential or bound in pairs to form meson-like excitations. Our results, supported by both exact analytical calculations and numerical simulations, suggest that confinement is a general mechanism underlying random state generation in broader settings with local measurements, including quantum circuits and chaotic dynamics.
\end{abstract}

\maketitle

\section{Introduction}
Randomness is a fundamental feature of quantum mechanics, manifested through measurement-induced collapse of quantum superpositions with probabilities dictated by the amplitude of the wave function. 
Beyond its foundational significance, quantum randomness plays a crucial role in modern quantum technologies~\cite{Preskill2018quantum,Morvan2024phase,Arute2019quantum}, from cryptographic protocols to benchmarking quantum devices~\cite{Gisin2002quantum,Liu2025certified}. However, generating and characterizing large-scale random quantum states remains a fundamental challenge.

A central notion in this context is that of quantum state designs~\cite{Gross2007evenly,Brando2016local,Harrow2009random}, which quantify how well an ensemble of states approximates the uniform probability measure on the Hilbert space (usually dubbed the Haar measure). Although random quantum circuits have been extensively studied as a means to achieve such designs~\cite{belkin2024approximatetdesignsgenericcircuit,mittal2023localrandomquantumcircuits,riddell2025quantumstatedesignsminimally}, the question of how randomness emerges in more constrained quantum architectures remains open. In particular, Tensor Network states~\cite{Collura_2024, Schollwock_2011}--ubiquitous in condensed matter physics~\cite{Orus_2019,Jordan_2008}, quantum simulations~\cite{Tindall_2024,Tindall_2025}, and holography~\cite{Hayden_2016, Vasseur_2019, Jahn_2021}-- offer a natural yet unexplored framework to investigate this question.

\begin{figure}[h!]
\centering
\includegraphics[width=\columnwidth]{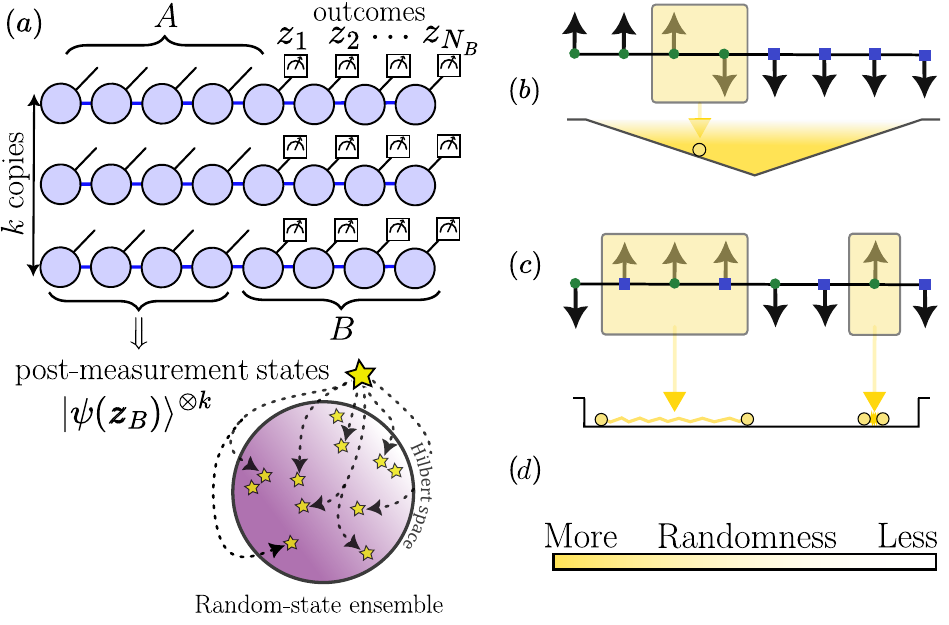}
\caption{Sketch of the work.
$(a)$ We consider a protocol where a random 
Matrix Product State (MPS) is prepared and subsequently subjected to projective measurements on the computational basis over a region $B$. The post measurement states in the complementary region $A$ are denoted $|\psi(\pmb{z}_B)\rangle$, where $\pmb{z}_B$ is the outcomes string. We investigate the randomness of these states—specifically, how uniformly they sample the Hilbert space. $(b)$ 
The quantifier of randomness, the frame potential, is mapped into the evaluation of the partition function of an effective spin chain, whose degrees of freedom are permutations. The study of this model reveals an underlying structure: in the scaling limit, domain walls emerge as quasiparticles confined by an effective potential. $(c)$ We also consider a different geometry, in which MPSs are generated by a glued shallow circuit and sites $A$ and $B$ are alternated. In this setting, confinement also emerges in the form of couples of domain walls bounded by a confining interaction.
$(d)$ In both cases the randomness of the projected ensemble is quantitatively related to the strength of the confinement.
}
\label{fig:sketch}
\end{figure}

In this work, we study the emergence of quantum randomness in random tensor networks subjected to local measurements~\cite{Choi2023preparing}. Specifically, we consider the ensemble of states obtained by projecting a random matrix product state (RMPS) or a random projected entangled pair state (PEPS) onto a measurement basis. The statistical properties of the resulting projected ensemble are analyzed using the frame potential, a standard tool for quantifying proximity to quantum designs. Through a mapping to a statistical mechanics model of interacting domain walls in replica space, we uncover a general mechanism: randomness emerges from a confinement effect. Here, confinement refers to an emergent potential that favors domain wall localization near subsystem boundaries or binding into meson-like pairs. This phenomenon echoes the confinement in quantum chromodynamics (QCD)~\cite{wilson,Gattringer2010} and condensed matter systems~\cite{Kormos2016,surace2020lattice,Zhang2024}, where quarks are bound by a linearly growing potential and never observed in isolation. In our setting, the combination of quantum measurements and the circuit's geometry binds the domain wall excitations, enhancing the uniformity of the ensemble, yet producing clear deviations from the Porter-Thomas distribution, the latter being expected only for fully random states~\cite{porter1956}.
Going beyond the realm of tensor network states, our findings suggest that such confinement is a generic mechanism for the emergence of quantum randomness, with implications for random quantum circuits, chaotic quantum dynamics, and entanglement structure in many-body systems. By elucidating this fundamental mechanism, our work provides a new perspective on the structure of quantum ensembles and the generation of randomness in quantum matter.

\section{Results}

\subsection{Quantum state design via measurements}
Nontrivial forms of quantum randomness emerge when multiple constituencies (e.g., qudits) interact via random quantum operations—as realized for instance by sufficiently deep quantum circuits with randomly chosen gates.
In a $N$ qudit quantum system described by a $D=d^N$ dimensional Hilbert space $\mathcal{H}$, random (pure) states in $\mathcal{H}$ can be generated by initializing the system in the product state $|0\rangle^{\otimes N}$ and applying a global unitary $U$ sampled from the Haar measure over the unitary group $\mathcal{U}(D)$. The resulting ensemble, $\Haar = \{ \ket{\psi} = U \ket{0}^{\otimes N}, \, U \sim \mathcal{U}(D) \}$,
is uniform in the Hilbert space and thus maximally random. However, this approach is extremely resource-intensive and quickly becomes infeasible even for moderately sized systems $N$.

Therefore, it is suitable to look for more efficient methods to generate a random state ensemble $\mathcal{E}$, and then quantify its randomness using appropriate measures. A key metric is the distance between $\mathcal{E}$ and the $\Haar$ ensemble, which can be rigorously characterized through the $k$-th frame potential. This is defined as the $k$-th moment of the overlap between two states in the ensemble, i.e.\ $\mathcal{F}^{(k)}_{\mathcal{E}} = \Ex_{\psi,\psi' \sim \mathcal{E} } [|\langle \psi | \psi' \rangle |^{2k}]$, where $\mathbb{E}_{\psi \sim \mathcal{E} }[...]$ denotes the average over states sampled from $\mathcal{E}$. The frame potential captures how well the ensemble $\mathcal{E}$ uniformly samples the Hilbert space, with increasing $k$ probing higher orders of statistical accuracy. When $\mathcal{F}^{(k)}_{\mathcal{E}} = \mathcal{F}^{(k)}_{\Haar}$, with $\mathcal{F}^{(k)}_{\Haar} \simeq D^{-k} k!$ for large Hilbert space dimension $D$, the ensemble $\mathcal{E}$ becomes \textit{statistically indistinguishable from $\Haar$ for any observable acting on at most $k$ identical copies of the system}; such an ensemble is denoted as a $k$-design~\cite{Mele2024,Fava2024designs}.

A resource-efficient framework for generating random ensembles from a single global state $|\psi\rangle$ uses projective measurements~\cite{Claeys2022emergent,Cotler2023emergent,Mark2024maximum,Mark2023benchmarking,Ippoliti2022solvable,Choi2023preparing,ho2022exact,shaw2024universalfluctuationsnoiselearning,ho2025nonlocality}. The system is partitioned into two subsystems: the subsystem of interest $A$, composed of $N_A$ qudits each of dimension $d_A$, and its complement $B$ ($N_B$ qudits of dimension $d_B$). Projective measurements on $B$ in the computational basis of multiple realizations of the same state $\ket{\psi}$ generate a stochastic ensemble of states on $A$, namely
\begin{equation}
\mathcal{P}[\ket{\psi}] = \{ |{\psi(\pmb{z}_B)}\rangle, \, \, p(\pmb{z}_B) \} \, ,
\end{equation}
where $\pmb{z}_B = (z_1, ... , z_{N_B})$, with $z_i \in \{0,1, ..., d_B - 1 \}$, labels the outcomes of the measurements in $B$, while $p(\pmb{z}_B)=\Tr[(\Id_A \otimes |\pmb{z}_B \rangle \langle \pmb{z}_B| ) |\psi \rangle \langle \psi|]$ are the associated measurement probabilities and $|{\psi(\pmb{z}_B)}\rangle = (\Id_A \otimes \langle \pmb{z}_B|) | \psi \rangle / \sqrt{p(\pmb{z}_B)}$ are the normalized post-measurement states in $A$. $\mathcal{P}[\ket{\psi}]$ is known as the \textit{projected ensemble}. Although its average density matrix on $A$ reproduces the conventional reduced density matrix, $\rho_{\mathcal{P}[\ket{\psi}]} = \mathbb{E}_{\phi \sim \mathcal{P}[\ket{\psi}]} [\ket{\phi}\bra{\phi}] = \Tr_B[\ket{\psi}\bra{\psi}] = \rho_A$, the projected ensemble contains additional structure beyond this first moment. 
This extra information enables refined studies of thermalization in subsystem $A$, where $B$ effectively serves as a `measurement-driven bath'. 
The information is encoded in the statistical fluctuations of the ensemble, and can be captured by the frame potential of $\mathcal{P}[\ket{\psi}]$~\cite{ippoliti2023dynamical,Chan_2024} 
\begin{align}\label{eq:frame_pot}
\begin{split}
\mathcal{F}^{(k)}_{\mathcal{P}[\ket{\psi}]} &= \sum_{\pmb{z}_B, \pmb{z}_B'} p(\pmb{z}_B) p(\pmb{z}_B') |\braket{\psi(\pmb{z}_B)|\psi(\pmb{z}_B')}|^{2k} = \\ &= \sum_{\pmb{z}_B, \pmb{z}_B'} p(\pmb{z}_B)^{1 - k} p(\pmb{z}_B')^{1 - k} |\braket{\tilde{\psi}(\pmb{z}_B)|\tilde{\psi}(\pmb{z}_B')}|^{2k} \, ,
\end{split}
\end{align}
where  $|{\tilde{\psi}(\pmb{z}_B)}\rangle = \Tr_B[(\Id_A \otimes \langle \pmb{z}_B|) \, | \psi \rangle ]$ are the unnormalized post-measurements states in $A$.

The condition that the projected ensemble forms a $k$-design, i.e.\ $\mathcal{F}^{(k)}_{\mathcal{P}[|\psi\rangle]} = \mathcal{F}^{(k)}_{\text{Haar}}$, carries deep implications.
It bridges thermalization and eigenstate typicality with notions from statistical mechanics, condensed matter theory, and high-energy physics, revealing how quantum complexity underlies foundational principles across multiple disciplines~\cite{pappalardi2022eigenstate,pappalardi2025full,lami2025quantum,Roberts_2017,brandao2021models}.

A central challenge lies in characterizing the projected ensemble emergent from a globally pure quantum state $\ket{\psi}$ \textit{prepared with limited resources}. In this context, Tensor Network states offer an ideal playground, as they represent
bounded-entanglement states and are both physically relevant and theoretically tractable. 
Here, we focus on the projected ensemble $\mathcal{P}[\ket{\psi}]$ generated from typical (random) one-dimensional Tensor Networks, i.e.\ Matrix Product States (MPS) $\ket{\psi}$. This setting provides a physically motivated and experimentally feasible approach to construct random many-body states on a subsystem.

\subsection{Random MPS and Glued circuits}
Matrix product states are a class of one-dimensional tensor networks with bounded entanglement. 
Mathematically, an MPS is specified by a set of $N$ tensors $M^{(i)}_{\alpha \beta}(z_i)$, where $\alpha,\beta\in\{1,2,\dots,\chi\}$ are indices associated with a bond (or auxiliary space) of dimension $\chi$, and $z_i\in \{0,1,\dots,d-1\}$ label the computational basis of the $i$-th qudit~\cite{Schollwock_2011,Collura_2024}. 
The wave function amplitudes $\psi(\pmb{z}) = \langle z_1,\dots,z_N|\psi\rangle$ are obtained by contracting these tensors along their auxiliary indices. 
This structure can be conveniently visualized using tensor network diagram, where blocks denote tensors, lines connecting blocks are contractions, and lines with open ends are free indices~\cite{Collura_2024}. For instance, a generic MPS is given by
\begin{equation}\label{eq:mps}
   \begin{tikzpicture}[baseline=(current bounding box.center), scale=1.]
   \definecolor{mycolor}{rgb}{0.82,0.82,1.}
   \pgfmathsetmacro{\lv}{0.5}
   \pgfmathsetmacro{\lu}{0.25}
   \pgfmathsetmacro{\luw}{0.25}
   \draw[ultra thick, blue] (1,\lv+\lu) -- (5,\lv+\lu);
   \foreach \x in {1,...,5}{
        \draw[thick, black] (\x,\lv+\lu) -- (\x,3*\lv);
        \draw[thick, fill=mycolor] (\x,\lv+\lu) circle (0.3);
        \node[scale=0.55] at (\x,\lv+\lu) {$M^{( \ifnum\x<4 
            \x 
        \else
            \ifnum\x=4
                ...
            \else
                N 
            \fi
        \fi)}$};
    }
   \end{tikzpicture} \, ,
\end{equation}
where vertical black lines denote the physical qudits and horizontal blue lines represent the auxiliary (bond) indices --- viewable as virtual qudits of dimension $\chi$. 
We examine two distinct protocols for constructing projected ensembles from random matrix product states. In both cases, a random quantum circuit arranged in a specific architecture prepares the MPS, and a subset of the qudits is subsequently measured in the computational basis to define an ensemble of states over the remaining degrees of freedom.

We anticipate that the two architectures give rise to projected ensembles with \textit{distinct emergent structures and randomness properties}. 
In particular, analyzing both protocols through the lens of the frame potential reveals two qualitatively different \textit{confinement mechanisms} that govern the buildup of randomness in the projected states. This gives rise to a radically different scaling of the corrections with respect to the Haar distribution in terms of the bond dimension $\chi$. 

\paragraph{Staircase circuit.} 
The first protocol is based on the standard sequential construction of (random) matrix product states through \textit{staircase circuits}~\cite{garnerone2010typicality,haferkamp2021emergent,Haag2023typical,lami2025quantum,cheng2024pseudoentanglementtensornetworks,sauliere2025universalityanticoncentrationchaoticquantum,Lami2025anticoncentration}. 
Concretely, random gates $U^{(i)} \sim \mathcal{U}(d\chi)$ are applied to a chain of qudits, each initialized in the state $|0\rangle$.
These unitaries entangle each physical site with the auxiliary space, progressively building the MPS structure.
After the full state is prepared, the subsystem $B$, consisting of the last $N_B$ qudits -- including the last auxiliary leg of dimension $\chi$ -- are measured in the computational basis. 
This projects the global state and defines a reduced ensemble on the unmeasured subsystem $A$, visually 
\begin{equation}\label{eq:rmps1}
\vcenter{\hbox{\includegraphics[width=0.80\linewidth]{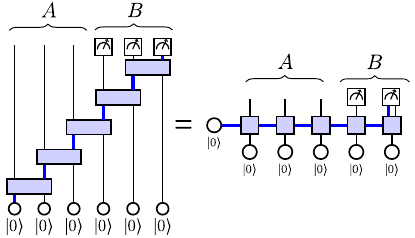}}}
\end{equation}
On the right-hand side of Eq.~\eqref{eq:rmps1}, we have rearranged the circuit to make the underlying MPS 
structure \eqref{eq:mps}  explicit. 
Note that, in this setting, all qudits in $A$ and $B$ are $d_A=d_B=d$ dimensional, except for the final site in $B$ which has dimension $d_B=\chi$.

\paragraph{Glued shallow circuits.}
The second protocol uses a shallow, two-layer circuit with a patched structure to build the random matrix product state~\cite{piroli2021quantum,malz2024preparation,Chen2024nishimori, Stephen_2024, Smith_2024, Sahay_2024, Zhang_2024}.
These so-called glued circuits have attracted significant attention within the quantum information community, as they offer a minimal and physically-relevant scheme for generating complex quantum states through local circuits. 
In particular, variants of the glued circuit architecture have been used in recent works to establish rigorous results on the preparation of approximate quantum state designs using low-depth random unitaries~\cite{laracuente2024approximateunitarykdesignsshallow,schuster2025randomunitariesextremelylow,magni2025anticoncentrationcliffordcircuitsbeyond,yada2025nonhaarrandomcircuitsform}. 
In this work, we give a quantitative prediction for a projected, measured-induced randomness, directly showing how advantageous this protocol is to create approximate quantum state designs. 

Concretely, this setup alternates between physical qudits of dimension $d$ and pairs of auxiliary qudits of dimension $\chi$, all initialized in the state $|0\rangle$. The preparation proceeds in two steps. 
In the first layer, gates $U^{(i)} \sim \mathcal{U}(d\chi^2)$ act to disjoint blocks consisting of one physical qudit and its two adjacent auxiliary qudits. 
These gates entangle the sites locally and can be applied in parallel across the chain.
In the second step, additional gates $U^{(i)} \sim \mathcal{U}(\chi^2)$ are applied between neighboring auxiliary qudits. 
These gates ``glue'' the blocks together by entangling adjacent virtual sites. 
Finally, all auxiliary qudits are measured in the computational basis, which projects the global state and defines the ensemble on the remaining (physical) sites.

The full process is visualized below 
\begin{equation}\label{eq:rmps2}
\vcenter{\hbox{\includegraphics[width=0.90\linewidth]{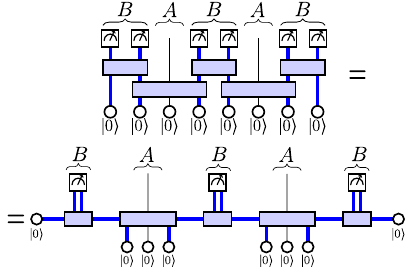}}}
\end{equation}

This architecture results in a matrix product state with an efficient and parallelizable preparation scheme. The projected ensemble is defined on the $N_A$ physical qudits of dimension $d_A=d$. In the second line of Eq.~\eqref{eq:rmps2}, we see that two adjacent auxiliary qudits can be combined into $N_B = N_A + 1$ $d_B$-dimensional local Hilbert spaces undergoing projective measurements, with $d_B = \chi^2$.

\subsection{Mapping to a statistical model of permutations}

We aim to compute the frame potential for the projected ensemble generated by the RMPS measurement protocols I and II.  
For $k > 1$, the negative exponents in the first two terms of Eq.~\eqref{eq:frame_pot} make analytical treatments problematic. We therefore employ the replica trick, defining the \textit{generalized frame potential} as
\begin{equation}\label{eq:frame_pot_1}
\mathcal{F}^{(k,n)}_{\mathcal{P}[\ket{\psi}]} \equiv \sum_{\pmb{z}_B, \pmb{z}_B'}p({\pmb{z}_B})^n  p({\pmb{z}_B'})^n |\braket{\tilde{\psi}(\pmb{z}_B)|\tilde{\psi}(\pmb{z}_B')}|^{2k}  \, .
\end{equation}
The original frame potential can be recovered setting $n = 1-k$, as $\mathcal{F}^{(k)}=\mathcal{F}^{(k,1-k)}$. 
For both technical advantage and to focus on general and state-independent features, we introduce an average over the global states $\ket{\psi}$ sampled over the RMPS ensemble, namely 
  $\mathcal{F}^{(k,n)} \equiv \Ex_{\psi}[\mathcal{F}^{(k,n)}_{\mathcal{P}[\ket{\psi}]}]$
where the average $\Ex_{\psi}$ is equivalent to the average over the circuit gates shown in Eqs.~\eqref{eq:rmps1}, \eqref{eq:rmps2}. 
For clarity, we stress that the frame potentials (\ref{eq:frame_pot}, \ref{eq:frame_pot_1}) involve the average over the states from the projected ensemble at fixed $\ket{\psi}$ and only afterwards the average over $\ket{\psi}$ is considered. 
By means of the replica trick,  Eq.~\eqref{eq:frame_pot_1} can be expressed  as the tensor product of $m=2(n+k)$ copies (or replicas) of the system, each including the ket state $\ket{\psi}$ and its conjugate bra $\bra{\psi}$ (see Eq.~\eqref{eq:frame_pot}). 
(Formally, this replica tensor network has a $\chi^{2m}$ dimensional auxiliary space.) 
The calculation of $\mathcal{F}^{(k,n)}$ reduces to contracting this replica tensor network with an appropriate product state, which encodes
the measurements in $B$ and the contractions between $\ket{\tilde{\psi}(z_B)}$ and $\ket{\tilde{\psi}(z_B')}$ in $A$. 
Then, the average over realizations of the circuits can be efficiently expressed with the toolbox of Weingarten calculus~\cite{Collins_2022, Mele2024}, which allows averaging the $m$ copies of each individual Haar gate $(U\otimes U^\ast)^{\otimes m}$. 
We leave the details of this calculation to Methods, but at this point it is useful to point out that, similarly to Wick's theorem for Gaussian averages, Haar's expectation value projects the $\chi^{2m}$-dimensional auxiliary space in each bond onto a reduced set of states, corresponding to all distinct pairings of bra and ket replicas~\cite{Nahum_2017,Nahum_2018,zhou2020entanglement,deluca2024universalityclassespurificationnonunitary}. As pairings can be identified with elements of $S_{m}$, the symmetric group with $m$ elements, we express $\mathcal{F}^{(k,n)}$ as the partition function of a suitable $1d$ Ising-like statistical mechanical model, whose degrees of freedom are permutations $\sigma \in S_{2m}$. 
As it is customary in statistical mechanics, this partition function can be conveniently expressed as a product of \textit{transfer matrices}, each encompassing replicas of a single qudit and averaging over the local MPS tensor. To express these matrices, it is convenient to define the transposition distance $\dist(\sigma,\sigma')$, which is the minimum number of elementary transpositions required to transform $\sigma$ into $\sigma'$. Furthermore, we introduce the Gram matrix $G_{\sigma,\sigma'}(q) = q^{m - \dist(\sigma,\sigma')}$ and the Weingarten matrix $W(q)$, which corresponds to the pseudoinverse of $G(q)$.
This leads to the expression
\begin{equation}
\label{eq:contraction_F}
    \mathcal{F}^{(k,n)} = 
     (v_L| \mathcal{T}_1 \mathcal{T}_2\ldots \mathcal{T}_{N-1}  |v_R)
\end{equation}
where $|v_L)$, $|v_R)$ are suitable boundary vectors specific to cases I and II, and encoding the distinction between the chain’s edge sites and the bulk; the  $\mathcal{T}_i$ are $m!\times m!$ matrices representing the weight associated to neighboring permutations on site $i, i+1$. Specifically,
\begin{align}
\label{eq:transfdef}
     [\mathcal{T}_i]_{\sigma' \sigma} &= 
     T_{\sigma, \sigma'}(\chi, d) \upsilon^{(i)}_{\sigma}(d) \, ,
\end{align}
where $T(\chi, d) = W(d\chi) G(\chi)$ accounts for the contraction in the auxiliary space and $\upsilon_\sigma^{(i)}(d)$ is the onsite contribution. Its form depends on whether the site $i \in A$ or $i \in B$. In the former, $\upsilon_\sigma^{(i)}(d) = 
\upsilon_\sigma^{(A)}:=
 d^{m - \dist(\sigma,\muA)}$. Here, $\muA \in S_m$ represents the permutation
\begin{equation}\label{eq:miozio}
 \sigma_A = \raisebox{0.7ex}{
\begin{tikzpicture}[baseline=(current  bounding  box.center),scale=0.4]
    \pgfmathsetmacro{\wv}{0.25} % Horizontal spacing
    \pgfmathsetmacro{\ll}{0.25}   % Size of the T rectangles
    \pgfmathsetmacro{\wr}{0.4}    % Line width T rectangles
    \pgfmathsetmacro{\lx}{0.5} % Horizontal spacing
    \pgfmathsetmacro{\ly}{2.8} % Vertical spacing
    \draw [line width=\wv mm, black](-4*\lx,\ly*0) -- (-4*\lx,\ly*1.0);
    \draw [line width=\wv mm, black] plot [smooth] coordinates {(-1.8*\lx, \ly*0) (-1.4*\lx, \ly*0.25) (-1*\lx, \ly*0.35) (0, \ly/2)  (1*\lx, \ly*0.65)  (1.4*\lx, \ly*0.75) (1.8*\lx, \ly*1.0)};
    \draw [line width=\wv mm, black] plot [smooth] coordinates {(-1.8*\lx,  \ly) (-1.4*\lx, \ly*0.75) (-1*\lx, \ly*0.65) (0, \ly/2)  (1*\lx, \ly*0.35)  (1.4*\lx, \ly*0.25) (1.8*\lx, \ly*0)};
    \draw [line width=\wv mm, black](4*\lx,\ly*0) -- (4*\lx,\ly*1.0);
    \node[scale=0.7] at (-4*\lx, -0.35) {$n$};
    \node[scale=0.7] at (4*\lx, -0.35) {$n$};
    \node[scale=0.7] at (-4*\lx, \ly+0.35) {$n$};
    \node[scale=0.7] at (4*\lx, \ly+0.35) {$n$};
    \node[scale=0.7] at (-1.8*\lx, -0.35) {$k$};
    \node[scale=0.7] at (1.8*\lx, -0.35) {$k$};
    \node[scale=0.7] at (-1.8*\lx, \ly+0.35) {$k$};
    \node[scale=0.7] at (1.8*\lx, \ly+0.35) {$k$};
\end{tikzpicture}} \, ,
\end{equation}
which characterizes the region $A$, as it accounts for the fact that the $2k$ central replicas are contracted among themselves to reproduce the scalar product 
between the post-measurements states
$|\braket{\tilde{\psi}(\pmb{z}_B)|\tilde{\psi}(\pmb{z}_B')}|^{2k}$. In contrast, in region $B$, when summing over the outcomes of the measures at site $i$, one must distinguish the $d$ cases where $z_i = z_i'$ from the $d(d-1)$ cases where $z_i \neq z_i'$. In the former in fact, the same MPS tensor $M^{(i)}(z_i)$ appears in all $m$ replicas, while in the latter $M^{(i)}(z_i)$ and $M^{(i)}(z_i')$ appear $m/2$ times each. This mechanism induces a preference for the $\sigma \in \Sfact$, 
where $\Sfact := S_{m/2} \times S_{m/2} = \{(\sigma_1, \sigma_2) \, ; \, \sigma_1,\sigma_2 \in S_{m/2}\}$ is the subset of factorized permutations; explicitly 
for $i \in B$, $\upsilon^{(i)}_{\sigma} = d^2 \pmb{1}_F(\pi) + d(1 -\pmb{1}_F(\pi))$, where 
$\pmb{1}_F$ is the characteristic function of 
the set $\Sfact$. 
With these definitions, 
setup I corresponds to choosing the region $A = \{1,\ldots, N_A\}$, while setting II $A = \{1,3,\ldots,N = 2N_A - 1\}$ (with minor differences, see Methods). 
Thanks to Eqs.~\eqref{eq:contraction_F}, the calculation of the generalized frame potential reduces to multiplication of matrices of size $m! \times m!$. Nonetheless, the replica limit requires the calculation of these quantities for arbitrary $k$ and $n$, which remains a hard task in general. However, the structure of the problem is clear: the $T_{\sigma,\sigma'}(\chi, d)$ matrix acts as a two-site interaction term, while $\upsilon^{(i)}_\sigma$ constitutes an on-site potential, with a different shape in the $A$ and $B$ regions. The structure of the transfer matrix largely simplifies at large $\chi$ and we will proceed to a systematic analysis of this limit in the next Section. Here, it is instructive to consider the leading order for $\chi \to \infty$. In this case, the Gram matrix reduces to multiple of the identity $G_{\sigma, \sigma'}(\chi) \stackrel{\chi \to \infty}{=} \chi^{m} \delta_{\sigma, \sigma'}$. The same happens to the Weingarten matrix $W(d \chi)$, being the pseudoinverse of $G(d\chi)$. It follows that $T_{\sigma,\sigma'}(\chi, d) \stackrel{\chi \to \infty}{=} d^{-m} \delta_{\sigma, \sigma'}$. This implies a strong ferromagnetic interaction, so that the same permutation is propagated unmodified through all the chain, picking up a weight from onsite potentials. Additionally, 
the boundary vector $|v_R)$ encodes for the fact that the rightmost site in $B$ has dimension $d_B = \chi$. For $\chi \to \infty$, this restricts the relevant permutations to factorized ones, leading to
\begin{equation}
\label{eq:Frame0}
 \mathcal{F}^{(k,n,0)} := 
\lim_{\chi \to \infty} \mathcal{F}^{(k,n)} = 
\frac{d^{-m N_A}}{d^{(m-2)(N_B - 1)}} \sum_{\sigma \in \Sfact}
[\upsilon_\sigma^{(A)}]^{N_A}
\;,
\end{equation}
where we used that $\upsilon^{(B)}_\sigma = d^2$ for each $\sigma \in \Sfact$.
Finally, at large $N_A$, only those permutations that minimize the distance to $\muA$ dominate. Regardless of $n$, there are precisely $k!$ of such permutations,
written as products of $k$ commuting transpositions $\{\tau_i\}$, i.e. $\sigma = \tau_{1} \tau_2 \ldots \tau_k \muA$, so that
$\dist(\sigma, \muA)=k$~\cite{ippoliti2023dynamical,Chan_2024}.
Thanks to this, and considering the replica limit $n \to 1-k$ ($m\to 2$), Eq.~\eqref{eq:Frame0} simplifies further, yielding the Haar frame potential for subsystem $A$, i.e.\ $\mathcal{F}^{(k,1-k,0)} = \mathcal{F}^{(k)}_{\Haar} = k! D_A^{-k}$. Consequently, for $\chi \to \infty$, the projected ensemble generates maximally random states in subsystem $A$.
Note that the $k!$ factor reflects the ground-state degeneracy associated with factorized permutations at minimal distance from $\sigma_A$ (see Methods for details). 

At any finite $\chi$, however, the ferromagnetic interaction remains finite and the off-diagonal terms in $T(\chi, \sigma)$ allow domain walls interfaces separating distinct permutations. The effect of these excitations will be addressed in the two following sections.

\begin{figure*}[t!]
\centering
\includegraphics[width=1.0\linewidth]{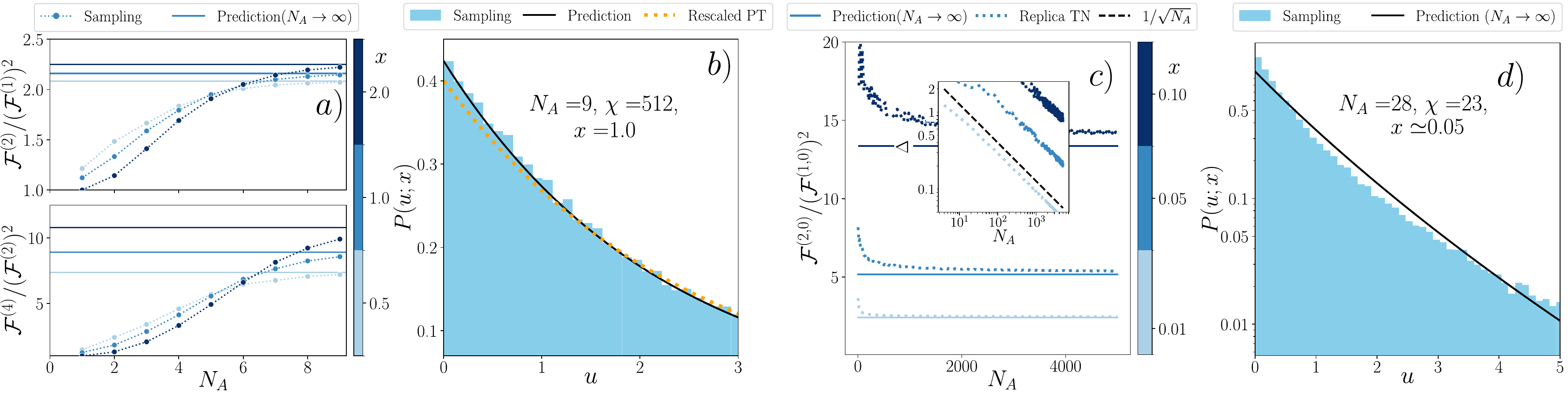}
\caption{\textbf{Projected ensemble - staircase RMPS (setup I)}. $a)$ Ratios of the moments $\mathcal{F}^{(k)}$ of the projected ensemble. Numerical data obtained with Born rule sampling of random MPS ($1500$ measurements per state, averaged over $100$ realizations) are compared with analytical predictions (horizontal lines) as function of the sub-system size $N_A$.  The MPS bond dimension scales as $\chi = 2^{N_A}/ x$, while the size of subsystem $B$ is $N_B = \lfloor N_A^{3/2} \rfloor$. $b)$ Distribution $P(u; x)$ of the overlaps $u$. We set $N_A = 9$, $\chi=512$, $x=D_A / \chi = 1.0$. The histogram of numerical samples obtained with Born rule sampling of random MPS (same samples as before) is compared with the analytical prediction for $P(u; x)$ (black line), and with the Porter-Thomas (PT) distribution rescaled by the expected mean (orange dotted line). The analytical prediction captures the behavior at $u=0$, while a simple rescaled PT does not.
\textbf{Projected ensemble - glued shallow circuit RMPS (setup II)}. $c)$ Ratios of the generalized frame potential $\mathcal{F}^{(k,n)}$ for $n=0$ (forced sampling of bitstrings). The MPS bond dimension scales here as $\chi = \lfloor \sqrt{N_A/x} \rfloor$. Results obtained through exact replica Tensor Network contraction (dotted lines) are compared with our analytical prediction, valid in the limit $N_A \to \infty$ (horizontal lines). To assess the convergence in $N_A$, we consider the difference between these two quantities $\Delta = \mathcal{F}^{(2,0)}/(\mathcal{F}^{(1,0)})^2 |_{\text{Replica TN}}-\mathcal{F}^{(2,0)}/(\mathcal{F}^{(1,0)})^2 |_{\text{Prediction}}$ (inset), which shows a  convergence with finite-size effects $\sim 1/\sqrt{N_A}$(black dotted line). $d)$ Distribution $P(u; x)$ of the overlaps $u$. We set $N_A = 28$, $\chi=23$, $x = 0.05 \simeq N_A / \chi^2$. The histogram of numerical samples obtained with Born rule sampling of random MPS is compared with the analytical prediction for $P(u; x)$ (black line). Given the numerically accessible small values of $N_A$, strong finite size effects prevent a full agreement in this case.}
\label{fig:plot1}
\end{figure*}

\subsection{Confinement mechanisms}

\subsubsection{Confinement for RMPS obtained through staircase circuits (setup I)
\label{sec:confiI}
}

To account for the formation of various domain walls, it is useful to expand the interaction matrix in powers of $1/\chi$ as 
\begin{equation}
\label{eq:Texp}
T_{\sigma', \sigma}(\chi, d) =d^{-m} (\delta_{\sigma', \sigma} +  \sum_{\alpha \geq 1} \adT^{(\alpha)}_{\sigma', \sigma} \, \chi^{-\alpha})
\end{equation}
where $\adT^{(\alpha)}$ are matrices encoding distinct hoppings between permutations. Specifically, $\adT^{(1)}=\frac{d-1}{d} \ad$, where $\ad$ is the adjacency matrix of $S_{m}$, i.e.\ the matrix that connects permutations at distance $1$ ($\ad_{\sigma', \sigma}=1$ iff $\dist(\sigma',\sigma)=1$, and $\ad_{\sigma', \sigma}=0$ otherwise). The adjacency matrix $\ad_{\sigma', \sigma}$ has recently appeared in replica calculations for the overlap distribution~\cite{christopoulos2025universaldistributionsoverlapsgeneric, Lami2025anticoncentration} and the purification dynamics under measurements~\cite{deluca2024universalityclassespurificationnonunitary}. In all those cases, a similar construction in terms of the transfer matrix (Eq.~\eqref{eq:transfdef}) was possible, but without any on-site potential term. Thus, excitations were simply domain walls (kinks) connecting different ferromagnetic ground states, with $1/\chi$ playing the role of fugacity. As domain walls could appear anywhere in the system, the relevant scaling regime in the thermodynamic limit required keeping fixed the ratio $N/\chi$. Here, instead, the presence of the onsite potential radically changes this scenario. To understand the mechanism, let us consider the expansion $\mathcal{F}^{(n,k)} = \sum_{\alpha \geq 0} \mathcal{F}^{(n,k,\alpha)}$,
where each $\mathcal{F}^{(n,k,\alpha)} = O(\chi^{-\alpha})$. While the first term, $\alpha=0$, was discussed above, we now analyze the first order, $\alpha=1$. This term
is obtained by replacing, in the product \eqref{eq:contraction_F}, a single $T(\chi, d)$ with the adjacency matrix $\adT^{(1)}$, while all the others are still multiple of the identity (as in the previous Section). Concretely, the matrix element $\adT^{(1)}_{\sigma', \sigma}$ on a certain bond $j$ is associated with a domain wall, so that all on-site potentials to the right are evaluated on $\sigma$, while those to the left in $\sigma'$, leading to
\begin{equation}
\label{eq:F1basic}
\mathcal{F}^{(n,k,1)} = \frac{d^{-m(N-1)}}{\chi}\sum_{\sigma, \sigma'} \sum_{j=1}^{N-1} \bigg[ \prod_{i=1}^j \upsilon_{\sigma'}^{(i)} \bigg] [\adT^{(1)}]_{\sigma',\sigma} \bigg[\prod_{i=j+1}^{N-1} \upsilon_{\sigma}^{(i)} \bigg] \;.
\end{equation}
In this sum, the rightmost spin in $B$ restricts $\sigma \in \Sfact$, as discussed above. We want to analyze Eq.~\eqref{eq:F1basic} in the regime $N_B \gg N_A \gg 1$ and we are therefore interested in how the factor $1/\chi$ can be offset by the possible choices of $\sigma'$. Since factorized permutations already maximize onsite contributions in $B$, to compensate for the $1/\chi$ factor the only possibility is for the permutation $\sigma'$ to be closer to $\muA$. Indeed, in this case starting with $\sigma = \tau_1 \ldots \tau_k \muA$ as one of the $k!$ ground states at distance $k$ from $\muA$, there are precisely $k$ choices $\tau \in  \{\tau_i\}_i$ that make $\dist(\sigma' = \tau \sigma, \muA) = k-1$. 
Comparing Eq.~\eqref{eq:F1basic} with \eqref{eq:Frame0}, this results in a factor $\upsilon_{\sigma'}^{(A)}/\upsilon_{\sigma}^{(A)}=d$ for each site in $A$ where such a $\sigma'$ replaces $\sigma$. 
Finally, the sum over $j$, i.e., the location of the domain wall, must be addressed. It is convenient to consider the possible positions of the domain wall relative to the interface between $A$ and $B$, i.e., $\ell = j - N_A$. First, if $\ell<0$, the domain wall is inside region $A$ and the gain due to $\sigma'$ replacing $\sigma$ equals $d^{N_A - |\ell|}$. Conversely, if $\ell>0$, the domain wall is inside region $B$ and the entire region $A$ is covered by $\sigma'$ resulting in a gain $d^{N_A}$. However, since $\sigma'$ cannot be factorized, there is a suppression factor $\upsilon_{\sigma'}^{(B)}/\upsilon_{\sigma}^{(B)} = d^{-1}$ for each site in $B$ where $\sigma'$ replaces $\sigma$. From this argument, we see that the natural scaling variable in the limit $N_B \gg N_A \gg 1$ and $\chi \to \infty$ is  
\begin{equation}\label{eq:scaling_1}
    x = \frac{D_A}{\chi} \frac{d-1}{d} = \frac{d^{N_A}}{\chi} \frac{d-1}{d} \, ,
\end{equation}  
where we inserted for convenience the factor $\frac{d-1}{d}$ (of order $1$), as it arises from the hopping matrix $\adT^{(1)}$ (see above).
We will denote as ``$\scaleq$'' equalities in the scaling limit at fixed $x$. Taking into account the two signs of $\ell$, one has
\begin{equation}
\label{eq:F1scalinglim}
\mathcal{F}^{(n,k,1)} \scaleq \mathcal{F}^{(n,k,0)} \, x k \sum_{\ell = -\infty}^{\infty} d^{- |\ell|} = \mathcal{F}^{(n,k,0)} \, x k \frac{d+1}{d-1}  \, .
\end{equation}
Note that in the scaling limit considered both $N_A,N_B \to \infty$, which allowed us to extend the geometric sum over $\ell$ to infinity. In practice, we can see Eq.~\eqref{eq:F1scalinglim} as the partition function for the position of the domain wall, in a confining potential of strength $\propto |\ell|$. It is worth dwelling on the interpretation of this result from the perspective of a quantum circuit with depth increasing with time $t$. In this case, the bond dimension grows exponentially with the depth $\chi \sim e^{v_{E} t}$, where $v_E$ is known as entanglement velocity~\cite{Nahum_2017}. Taking the replica limit $n \to 1-k$, the expansion of the frame potential to first order in $1/\chi$ reads
\begin{equation}
    \mathcal{F}^{(k)} \sim k! d^{-N_A}\left(1 + x k \frac{d+1}{d-1}\right) =
    k! (d^{-N_A} +  \frac{k(d+1)}{d} e^{-v_E t}) \;.
\end{equation}
where in the last expression we wrote the explicit form of $x$ \eqref{eq:scaling_1}. The latter expression has a natural interpretation in terms of picture membranes that emerged in the context of random circuits~\cite{Nahum_2017}: the two terms in parentheses correspond, respectively, to configurations in which the membrane exits from the left edge of the system (with a cost $O(N_A)$) and those in which it instead exits from the lower edge (with a cost $O(t)$)~\cite{adam, Chan_2024}. Thus, the description by random tensor network nicely matches with the one of circuits.

\begin{figure}[ht]
%\vcenter{\hbox{
\includegraphics[width=1.\linewidth]{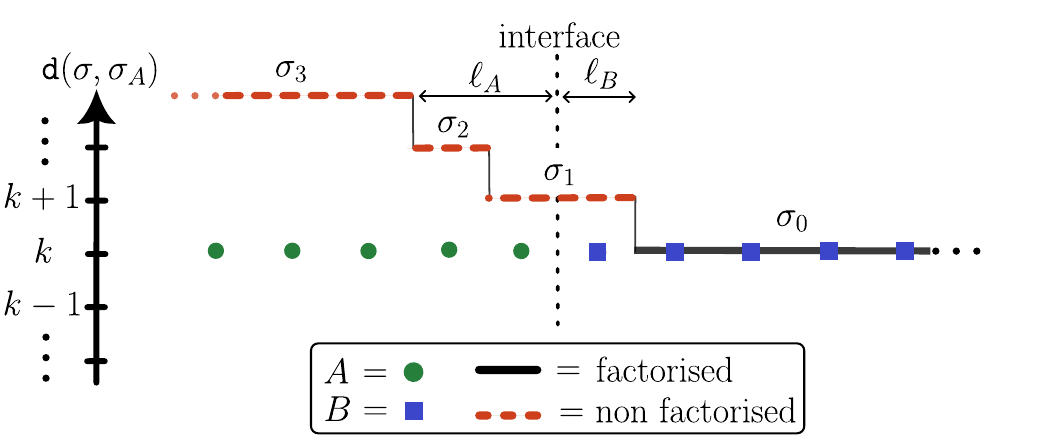}
%}}
\caption{A graphical representation of domain walls excitations and confinement for staircase RMPS (setup I).
We consider dominant terms in the scaling limit in the expansion of the frame potential at order $\alpha$ in powers of $\chi^{-1}$, i.e.\ $F^{(n,k,\alpha)}$ ($\alpha=3$ in this case). Each domain wall modifies the permutation by bringing it closer to $\muA$.
\label{fig:eccitazioni_1}
}
\end{figure}

At this point, we are in a position to discuss subsequent orders as well. These are controlled by insertions of the matrices $\adT^{(\beta)}$ for $\beta>1$, but for now let us analyze $\beta=1$. Then, $F^{(n,k,\alpha)}$ will involve $\alpha$ elementary domain walls (transpositions) associated with as many insertions of the matrix $\adT^{(1)}$. This induces the presence of $\alpha+1$ domains: $\sigma_{\alpha},\sigma_{\alpha-1},\ldots,\sigma_1, \sigma_{0} = \sigma$ with $\sigma \in \Sfact$. From the scaling limit \eqref{eq:scaling_1}, it follows that to compensate for the factor $\chi^{\alpha}$, the leftmost domain must lie at a distance $\dist(\sigma_{\alpha}, \muA) = k - \alpha$ and occupy $N_A - O(1)$ sites.
This implies a rigid structure:  proceeding from right to left, each domain wall brings the permutation progressively closer to $\muA$, i.e., $\dist(\sigma_a, \muA) = k - a$, see Fig.~\ref{fig:eccitazioni_1}. 
Given a ground state $\sigma_0 = \tau_1\ldots \tau_k \muA$, the $\alpha$ domain walls can be associated with any of the $k!/(k-\alpha)!$ ways of selecting $\alpha$ elements from $\{\tau_i\}$ considering their ordering. Additionally, we briefly discuss the positioning of the domain walls $\ell_{\alpha}< \ell_{\alpha-1}< \ldots< \ell_1$. As clarified earlier, domain walls in $A$ ($\ell_i < 0$) produce progressive approaches to $\muA$. In contrast, in $B$, any $\sigma_a \neq \sigma$ will not be factorized and  encounters the same $1/d$ suppression factor. This induces an asymmetric potential in the two regions
\begin{equation}
    V(\ell_1,\ldots, \ell_{\alpha}) = V_A(\ell_1,\ldots, \ell_{\alpha}) + V_B(\ell_1,\ldots, \ell_{\alpha}) \, ,
\end{equation}
and on the basis of the previous discussion, the exact form of $V_A, V_B$ reads
\begin{align}\label{eq:VA_e_VB}
    &V_A(\ell_1,\ldots, \ell_{\alpha}) = \sum_{i=1}^{\alpha} \max(-\ell_i, 0) \\
    &V_B(\ell_1,\ldots, \ell_{\alpha}) = \max(\ell_1,\ldots, \ell_{\alpha}, 0) \, \;.
\end{align}
Finally, we deal with the subleading corrections $\adT^{(\beta)}$ for $\beta>1$. They are associates with cases where two or more elementary domain walls are located on the same bond, i.e.\ $\ell_i = \ell_{i+1} = \ldots = \ell_{i+\beta-1}$. The $\adT^{(\beta)}$ matrices contain the information of how the domain walls interact. This aspect is a major difference with the scaling limits analyzed in Refs.~\cite{christopoulos2025universaldistributionsoverlapsgeneric, Lami2025anticoncentration}: there, the scaling limit $N/\chi = \text{fixed}  $ implied that the domain walls were indeed diluted, so that all $\adT^{(\beta>1)}$ were in fact irrelevant, giving rise to a mechanism for universality. Here instead, they remain relevant. However, for the Haar unitary RMPS we consider here, it is still possible to treat the interactions exactly, since domain walls are associated with transpositions that commute. Specifically, putting 
$p$ commuting transposition on the same bond, one has
simply $\mathsf{T}_{\tau_1\ldots\tau_p, \mathbf{1}}^{(p)} = ((d-1)/d)^p$ which is conveniently absorbed in the definition \eqref{eq:scaling_1} of $x$ (see Methods).
We obtain simply the final expression in the scaling limit \eqref{eq:scaling_1}
\begin{equation}
    \mathcal{F}^{(n,k,\alpha)} \scaleq \mathcal{F}^{(n,k,0)} \binom{k}{\alpha} x^{\alpha} \sum_{\ell_1,\ldots, \ell_{\alpha}=-\infty}^\infty d^{- V(\ell_1,\ldots, \ell_{\alpha})}
\end{equation}
Summing up all the contributions, we obtain the following expression for the frame potential in the replica limit  
\begin{equation}\label{eq:result_1}
   \mathcal{F}^{(k)}(x) = \mathcal{F}^{(k)}_{\Haar} \frac{d-1}{d} \sum_{j = 0}^{+\infty} d^{-j} \left(1 + x \frac{d}{d-1} + x j  \right)^k \, ,
\end{equation}
where the sum over the discrete variable $j$ is reminiscent of the fact that the confining length $\ell_A, \ell_B$ can only take integer values.  
The expression Eq.~\eqref{eq:result_1} recovers previous results~\cite{ippoliti2023dynamical,Chan_2024} in the limit of large qudit dimension $d \gg 1$, where the sum reduces to $\mathcal{F}^{(k)}(x)  \approx \mathcal{F}^{(k)}_{\Haar} (1 + x)^k$. However, our expression is valid for generic $d$, and it remarkably shows how finite physical dimension gives rise to discrete fluctuation of the interface. In Fig.~\ref{fig:plot1}, we compare our analytical predictions for the frame potential at small 
$k$ with numerical results obtained by sampling random MPS realizations. The convergence to our theoretical prediction is clear, with excellent agreement even for moderately small values of $N_A$. Furthermore, our analysis remains valid as long as the size of region $B$ scales polynomially faster than $N_A$, specifically $N_B \sim N_A^{\nu}$ with exponent $\nu>1$. This fact indicates that a relatively small bath is sufficient in order to induce randomness in $A$ up to a finite precision. 

\subsubsection{Confinement for RMPS obtained through glued shallow circuits (setup II)}

\begin{figure}[ht]
\includegraphics[width=1.\linewidth]{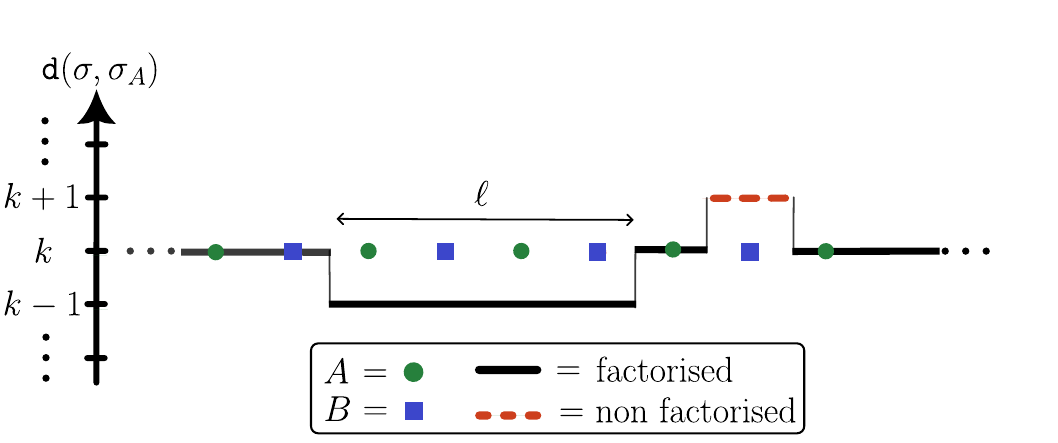}
\caption{A graphical representation of domain walls excitations and confinement for glued RMPS (setup II).
We consider dominant terms in the scaling limit in the expansion of the frame potential at order $\alpha$ in powers of $\chi^{-1}$, i.e.\ $F^{(n,k,\alpha)}$ ($\alpha=4$ in this case). 
\label{fig:eccitazioni_2}
}
\end{figure}

As in the previous case, excitations can appear at cost $1/\chi$ per domain wall. However, since sites $A$ and $B$ are now alternating and the local dimension on $B$ sites is $\chi^2$, the nature of excitations changes significantly.
As before, a domain wall may be created to bring the permutation closer to $\muA$ on a site $A$. However, this process necessarily creates a non-factorized permutation. Because such permutation incur an energy cost of $1/\chi^2$ at every $B$ site, 
the resulting excitations must be annihilated by a compensating domain wall that restores the original factorized permutation without including any site in $B$ (see dashed red line in Fig.~\ref{fig:eccitazioni_2}). Overall, these kink-antikink excitations therefore occur at a cost of $1/\chi^2$, and since they can appear on all $A$ sites, their natural scaling variable is  
\begin{equation}\label{eq:scaling_2}
    x = \frac{N_A}{\chi^2} \, .
\end{equation}
Consequently, we will consider the scaling limit where both $N_A$ and $\chi$ are large while $x$ remains fixed. In this limit, additional excitations with the same scaling behavior emerge. For instance, at any point of the chain the current permutation can `jump' to another factorized permutation at distance $\dist(\sigma, \muA)=k+1$ via a domain wall, persist in this configuration over $\ell$ sites, and subsequently revert to the original state. Crucially, the domain wall pairs associated with these excitations experience a confining interaction potential, as their contribution is suppressed by a factor $1/d$ for each site $A$ involved, a quantity scaling linearly with $\ell$ (see black thick line in Fig.~\ref{fig:eccitazioni_2}).
Thus, these excitations are meson-like with a $O(1)$ extension and can be created on all sites, giving a scaling consistent with Eq.~\eqref{eq:scaling_2}.

In Figure~\ref{fig:eccitazioni_2}, $\ell$ is the confinement length and the amplitude associated to a finite $\ell$ scales as $d^{-\ell}$. 
Our analysis of setup II's statistical model reveals additional excitation types obeying the same scaling (Eq.~\eqref{eq:scaling_2}), which we omit here for brevity. 
Summing up all these contributions we obtain a rather involved expression for $\mathcal{F}^{(k,n)}(x)$, which we omit here. The replica limit in this case is non-trivial, as the expression exhibits an explicit dependence on the replica index $n$. Taking the limit $n \rightarrow 1-k$, the frame potential reduces to
\begin{equation}\label{eq:result_2}
   \mathcal{F}^{(k)}(x) = \mathcal{F}^{(k)}_{\Haar} \exp \left(x k  (  d-1-1/d ) +  x k^2 \right) \, .
\end{equation}
In Fig.~\ref{fig:plot1}, we consider the frame potential for the first values of $k$. To access large systems' sizes, we plot the case with $n=0$, which corresponds to unnormalized states without Born rule, where eq. \eqref{eq:contraction_F} can be numerically evaluated. In this case, we perform exact numerical simulations for large $N_A$ by evaluating the matrices product in Eq.~\eqref{eq:contraction_F} (i.e.\ the so-called replica tensor network method). The plot shows that the approach to our asymptotic expression is rather slow, with corrections of order $N_A^{-1/2}$ (see inset). This fact makes a direct comparison with numerical data for the Born rule scenario ($n=1-k$) hard, as exact classical simulations of quantum systems are inherently limited to small sizes.

\subsection{The overlaps distribution}

The results we derived for the frame potential can be used to obtain exact analytical expressions for the probability distribution of the (rescaled) overlaps between different states in the projected ensemble, namely $D_A |\langle \psi(\pmb{z}_B)|\psi(\pmb{z}_B')\rangle|^2$ between states on $A$ conditioned on distinct measurement outcomes $\pmb{z}_B,\pmb{z}_B'$ in $B$. Specifically, we define
\begin{equation}\label{eq:prob_distr}
    P(u) \equiv \Ex_{\psi} \bigg[ \sum_{\substack{\pmb{z}_B, \\ \pmb{z}_B'}} p(\pmb{z}_B) p(\pmb{z}_B') \, \delta \left(u - D_A |\langle \psi(\pmb{z}_B)|\psi(\pmb{z}_B')\rangle|^2 \right) \bigg]  \, .
\end{equation}
Given that the $k$-th moment of $P(u; x)$ coincide with $\mathcal{F}^{(k)}$, $P$ is the distribution whose moments are the frame potentials. For global Haar states ($\psi \sim \Haar$), the distribution takes a universal form known as the Porter-Thomas distribution, $P_0(u)=e^{- u}$. For the projected ensemble of RMPS, within the setup I, we instead find 
\begin{equation}\label{eq:distributionI}
     P(u; x) = \frac{d-1}{d} \sum_{j=0}^{\infty} e^{- \frac{u}{z_{j}}} z_{j}^{-1}  d^{-j} \, ,
\end{equation}
where $z_{j} = 1 + x \frac{d}{d-1} + x j$, and where the notation explicitly reflects the parametric dependence of the distribution on the scaling variable $x$. Clearly, in the limit of large bond dimension $\chi \gg D_A$, we recover the Porter-Thomas distribution: $\lim_{x \to 0^+} P(u; x) = P_0(u)$. However, at any finite value of $x$, the distribution is non-trivial. 

In the limit of large $d$, the distribution converges to a simple rescaled Porter-Thomas, as well shown in Fig.~\ref{fig:plot1}. In fact, the finite local physical dimension gives important fluctuations around the interface, manifested in the discrete nature of the distribution \eqref{eq:distributionI}. 

In the setup II, we instead find 
\begin{equation}
    P(u;x) = \int_{-\infty}^{+\infty} \frac{dw}{\sqrt{2 \pi}} e^{-\frac{w^2}{2}-  w \sigma - \mu}
    \exp \big[ - u e^{-\sigma w - \mu}] \, ,
\end{equation}
where $\mu=x \frac{d^2 - d - 1}{d}$, $\sigma=\sqrt{2x}$. This distribution arises from a suitable convolution of a Porther-Thomas distribution (first factor in Eq.~\eqref{eq:result_2}), and a log-normal distribution with mean $\mu$ and variance $\sigma^2$ (second term in Eq.~\eqref{eq:result_2}), which, quite remarkably, has recently been shown to also appear universally in the inverse participation ratio of chaotic circuits ~\cite{Lami2025anticoncentration,christopoulos2025universaldistributionsoverlapsgeneric,sauliere2025universalityanticoncentrationchaoticquantum}. As specified in the previous section, large finite systems' size corrections make the comparison with numerical data at finite $N_A$ arduous, even if the scaling in $x$ and the general shape are well captured at such small sizes. 

\subsection{Generalization to higher-dimensional geometries}

\begin{figure}[ht]
\includegraphics[width=0.7\linewidth]{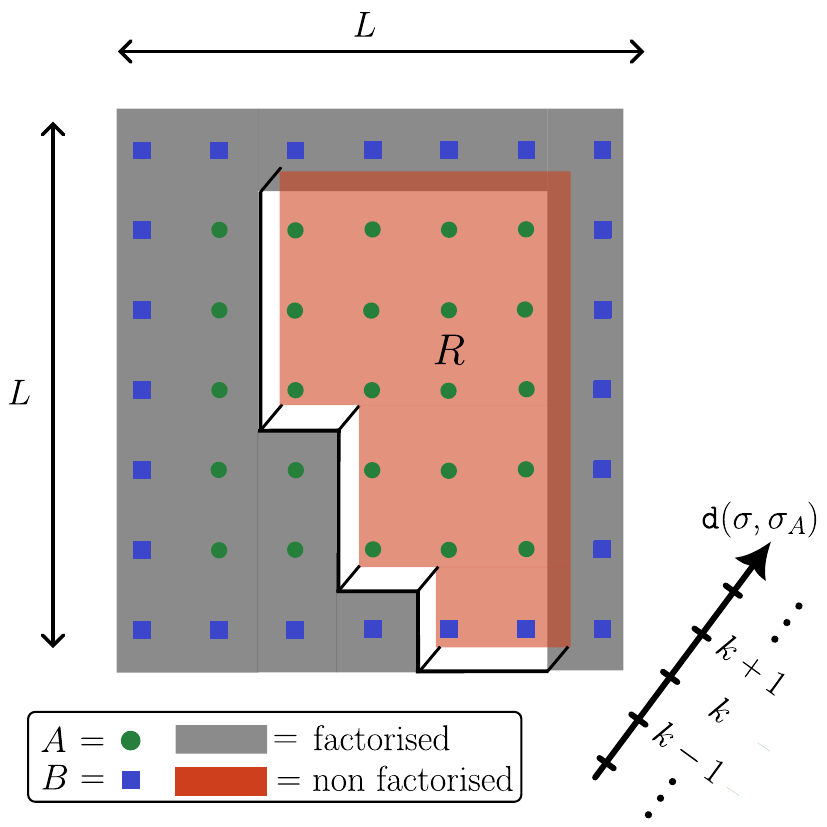}
\caption{A graphical representation of domain walls excitations for the staircase RMPS (setup I), in two-dimensional case.
\label{fig:eccitazioni_2D}
}
\end{figure}

Both architectures admit a natural two-dimensional generalization on an $L \times L$ square lattice of $N=L^2$ total sites. 
The staircase circuit architecture (setup I) can be extended to generate a 2D (random) isometric tensor network, specifically an isometric PEPS~\cite{Lami2025anticoncentration,zaletel2020isometric,jordan2008classical}. %,PhysRevA.105.022611}. 
In this setting, the system is partitioned into two distinct but topologically connected regions, $A$ and $B$, with measurements performed in the computational basis on one of them.
Although the analytical treatment becomes more involved in two dimensions, the essential features of randomness generation via confinement remain qualitatively robust. 
In particular, creating a non-factorized permutation excitation over a region $R\subset A$ incurs an energy cost scaling as $\sim d^{|R|}/\chi^{|\partial R|}$. Here, $\chi^{|\partial R|}$ is the domain wall energy cost along the boundary $\partial R$ of $R$, and $d^{|R|}$ reflects the energy gain from moving permutations closer to $\muA$. 
This structure favors extensive excitations that span large portions $R$ of region $A$, while localized defects are exponentially suppressed. 
This suppression is overt in the scaling limit $\chi, N_A, L \to \infty$, with $x \sim  D_A/\chi^L$ fixed. 
Moreover, extended excitations may also involve pars of subsystem $B$, experiencing a similar confinement potential as in the one-dimensional case, see Fig.~\ref{fig:eccitazioni_2D} for a sketch. 

Setup $II$ also admits a natural extension to two dimensions. Here, unentangled $A$ blocks can be arranged on a square lattice and prepared in parallel, with subsequent gates and measurements used to ``glue`` them together. 
The resulting structure closely mirrors the one-dimensional cases: local non-factorized domain walls can be introduced by inserting a transposition on each of the four bonds surrounding a site in region $A$, at an energy cost of $(1/\chi^2)^4=1/\chi^8$. This yields a defect density scaling as $x\sim N_A/\chi^8$. 

\section{Discussion}
In this work, we explore the emergence of quantum randomness from local measurements on tensor network states generated by suitable quantum circuits and characterized by a finite bond dimension $\chi$.
Focusing on the ensembles formed by projectively measuring a portion of the system, which effectively acts as a bath to the complementary, we demonstrate that the statistical properties of the unmeasured subsystem are well described by a domain wall statistical model with confining mechanisms. 
We analyzed two physically motivated architectures, both compatible with near-term quantum devices. 
The first setup (I), uses a staircase circuit to generate a random matrix product state, which is measured on the $N_B$ rightmost sites. 
In the second setup (II), reminiscent of recently studied glued circuits~\cite{schuster2025randomunitariesextremelylow}, an ensemble of random MPS, forming an approximated state design with only $\chi \sim N^2$, is efficiently prepared by first creating unentangled blocks in parallel and then entangling them via unitaries and measurements in the $\chi^2$-dimensional auxiliary space.

In both cases, we find that confinement plays a central role in shaping the projected ensemble. In setup I, domain walls, representing interfaces between different replica configurations, are energetically pinned to the interface with the measured region. The existence of a collective confined state of domain wall can be seen as the counterpart in tensor networks of the membrane observed in random unitary circuits~\cite{adam, Chan_2024}. In setup II, instead, domain walls are bound into meson-like pairs that remain confined throughout the bulk. This distinction gives rise to different types of emergent randomness.

A key outcome of our analysis is the identification of a scaling regime where the projected ensemble admits exact analytical treatment. In this regime, the bond dimension $\chi$ scales exponentially (setup~I) or, quite remarkably, polynomial (setup~II) with the subsystem size, making the second proposed protocol an ideal setting for engineering approximated quantum state designs. 
The scaling limit yields exact analytical expressions, verified numerically whenever possible, for the frame potential and overlap distribution of the projected ensemble. 
Both quantities take on universal forms that remain valid far beyond infinite-depth circuits, offering directly testable predictions and implementable state design protocols for current digital quantum devices. 

Interestingly, in the staircase architecture, the distribution of measurement outcomes exhibits discrete fluctuations that modulate the Porter-Thomas law, revealing a new structure in the interplay between quantum measurements and randomness. In contrast, the glued circuit provides a remarkably efficient pathway to generate approximate $k$-designs using only a polynomial bond dimension, offering a scalable route to structured quantum randomness.

Altogether, our results reveal confinement, ubiquitous in high-energy physics and in condensed matter theory, as the underlying mechanism governing the generation of randomness in quantum circuits. This work opens new avenues for understanding the structure of projected ensembles and suggests that local measurements in shallow architectures can serve as powerful tools for quantum information processing, benchmarking, and state design.

\newpage

\section{Methods}
We aim to compute the frame potential of the projected ensemble generated by a random matrix product state (RMPS), averaged over circuit realizations. Physically, all gates in Eqs.~\eqref{eq:rmps1} and~\eqref{eq:rmps2} are unitary and drawn from the Haar measure over the unitary group. To ease the explanation, however, we also consider a simplified ensemble in which each gate is replaced by a random matrix with i.i.d. complex Gaussian entries of zero mean and variance $\varsigma^2$. We refer to this as the $\Gauss $ ensemble, and denote both cases using the unified symbol  $\mathcal{E} \in \{\Haar, \Gauss\}$. As we will show, the key features of the projected ensemble remain qualitatively similar under either choice.
Methodologically, we employ Weingarten calculus within the Choi-Jamiolkowski representation, in which operators $O$ acting on a $q$-dimensional Hilbert space are mapped to states $O \mapsto |O\rangle\!\rangle$. 
Within this notation $U^\dagger O U$ maps into $(U^*\otimes U )|O\rrangle$, and the Hilbert-Schmidt inner product of two operators $O$ and $V$ corresponds to $\mathrm{tr}(O^\dagger V) = \langle\!\langle O|V\rangle\!\rangle$ (in some cases we add a subscript $\langle\!\langle O|V\rangle\!\rangle_q$ to emphasize the Hilbert space dimension).

\subsection{Haar and Gaussian averages} 
Our computations require evaluating the Haar moments $\Ex_{U \sim \Haar(q)}[(U^{*}  \otimes U)^{\otimes m}]$. These can be expressed in terms of permutation operators acting on $m$ copies, or replicas, of the Hilbert space~\cite{Collins_2022}. 
We will represent them graphically as the three legs tensors
% \begin{equation}
$\vcenter{\hbox{\begin{tikzpicture}[baseline=(current  bounding  box.center), scale=0.4]
\tikzset{snake it/.style={decorate, decoration=snake}}
\draw[ultra thick, black] (0.5,0.6) -- (-0.5,0.6);
\draw[ultra thick, black] (0.5,0.4) -- (-0.5,0.4);
\draw [ultra thick, orange, snake it] (0.5,0.5) -- (2.0,0.5);
\draw[ultra thick, fill=white] (0,0) rectangle (1.,1.);
\end{tikzpicture}}},$
% \end{equation}
where the wavy orange line $\vcenter{\hbox{\begin{tikzpicture}[baseline=(current  bounding  box.center), scale=0.5]
\tikzset{snake it/.style={decorate, decoration=snake}}
\draw [ultra thick, orange, snake it] (0.5,0.5) -- (2.0,0.5);
\end{tikzpicture}}}$ is the permutation index $\sigma\in S_m$, and each $\vcenter{\hbox{\begin{tikzpicture}[baseline=(current  bounding  box.center), scale=0.45]
\tikzset{snake it/.style={decorate, decoration=snake}}
\draw [ultra thick, black] (0.5,0.5) -- (2.0,0.5);
\end{tikzpicture}}}$ leg represents the $m$-th tensor product of the physical space, thus of dimension $q^m$. 
The central formula of Weingarten calculus reads~\cite{Collins_2022, Mele2024}  
\begin{equation}\label{eq:haark}
\begin{split}
    \Ex_{U \sim \Haar}&[(U^{*}  \otimes U)^{\otimes m}] =\sum_{\sigma, \pi \in S_m} \Wg_{\sigma,\pi}{(q)} |\sigma\rrangle  \llangle \pi| \, \\
    &=\raisebox{2.1ex}{\begin{tikzpicture}[baseline=(current  bounding  box.center), scale=0.55]
\tikzset{snake it/.style={decorate, decoration=snake}}
\pgfmathsetmacro{\ll}{3.5}
\draw[ultra thick, black] (0.5,0.6) -- (-0.5,0.6);
\draw[ultra thick, black] (0.5,0.4) -- (-0.5,0.4);
\draw[ultra thick, black] (\ll+0.5,0.6) -- (\ll+1.5,0.6);
\draw[ultra thick, black] (\ll+0.5,0.4) -- (\ll+1.5,0.4);
\draw [ultra thick, orange, snake it] (0.5,0.5) -- (4.0,0.5);
\draw[ultra thick, fill=white] (0,0) rectangle (1.,1.);
\draw[ultra thick, fill=white] (\ll,0) rectangle (\ll+1,1.);
\node[draw, diamond, fill=orange, ultra thick] (W) at (\ll/2+0.5, 0.5) {};
\node[scale=1.] at (\ll/2+0.5, 1.5) {$W(q)$};
\end{tikzpicture}}\;,
\end{split}
\end{equation}
where $W(q)$ is the Weingarten matrix $\Wg_{\sigma,\pi}{(q)}$~\cite{Collins_2022}, i.e., the inverse of the Gram matrix $G_{\sigma,\pi}=\llangle \sigma | \pi \rrangle_q=q^{m - \dist(\sigma, \pi)}$, graphically
\begin{equation}
\vcenter{\hbox{
\raisebox{2.5ex}{
\begin{tikzpicture}[baseline=(current  bounding  box.center), scale=0.8]
\tikzset{snake it/.style={decorate, decoration=snake}}
\pgfmathsetmacro{\ll}{1.5}
\draw [ultra thick, orange, snake it] (\ll/2-1,0.5) -- (\ll/2+1,0.5);
\node[draw, rectangle, fill=blue, ultra thick] at (\ll/2, 0.55) {};
\node[scale=1.] at (\ll/2, 1.) {$G$};
\end{tikzpicture} }
\scalebox{1.0}{$=$}
\raisebox{0.9ex}{
\begin{tikzpicture}[baseline=(current  bounding  box.center), scale=0.45]
\pgfmathsetmacro{\ll}{2.5}
\tikzset{snake it/.style={decorate, decoration=snake}}
\draw [ultra thick, orange, snake it] (0.5,0.5) -- (-1.,0.5);
\draw [ultra thick, orange, snake it] (\ll+0.5,0.5) -- (\ll+2.,0.5);
\draw[ultra thick, black] (0.5,0.6) -- (\ll+0.5,0.6);
\draw[ultra thick, black] (0.5,0.4) -- (\ll+0.5,0.4);
\draw[ultra thick, fill=white] (0.1,0) rectangle (1.1,1.);
\draw[ultra thick, fill=white] (\ll-0.1,0) rectangle (\ll+1.-0.1,1.);
\end{tikzpicture} \;.
}}}
\end{equation}
We consider also matrices with i.i.d.\ complex Gaussian entries, with zero mean and variance $\varsigma^2$. 
The Gaussian case can always be recovered from the Haar ensemble formulas, replacing 
$\Wg_{\sigma,\pi}{(q)} \to \delta_{\sigma \pi} \varsigma^{2m}$, i.e, 
\begin{equation}\label{eq:diagonal_Wg}
\raisebox{2.5ex}{
\begin{tikzpicture}[baseline=(current bounding box.center), scale=1.]
\tikzset{snake it/.style={decorate, decoration=snake}}
\pgfmathsetmacro{\ll}{1.5}
\draw [ultra thick, orange, snake it] (\ll/2-1,0.5) -- (\ll/2+1,0.5);
\node[draw, diamond, fill=orange, ultra thick] at (\ll/2, 0.5) {};
\node[scale=1.] at (\ll/2, 1.) {$W(q)$};
\end{tikzpicture}
}
\scalebox{2.0}{$=$}
\varsigma^{2m}
\raisebox{0.9ex}{
\begin{tikzpicture}[baseline=(current bounding box.center), scale=0.7]
\tikzset{snake it/.style={decorate, decoration=snake}}
\draw [ultra thick, orange, snake it] (-1.5,0.5) -- (+1.5,0.5);
\end{tikzpicture}
} \, ,
\end{equation}
leading to the Gaussian version of  Weingarten formula (Wicks' theorem)
\begin{equation}\label{eq:gauss_average}
\begin{split}
  \Ex_{A \sim \Gauss}&[(A^{*}  \otimes A)^{\otimes m}] \simeq \varsigma^{2m}  \sum_{\pi\in S_m} |\pi \rrangle  \llangle \pi| \, \\
    &= \varsigma^{2m} \, \raisebox{0.8ex}{\begin{tikzpicture}[baseline=(current  bounding  box.center), scale=0.55]
\tikzset{snake it/.style={decorate, decoration=snake}}
\pgfmathsetmacro{\ll}{3.5}
\draw[ultra thick, black] (0.5,0.6) -- (-0.5,0.6);
\draw[ultra thick, black] (0.5,0.4) -- (-0.5,0.4);
\draw[ultra thick, black] (\ll+0.5,0.6) -- (\ll+1.5,0.6);
\draw[ultra thick, black] (\ll+0.5,0.4) -- (\ll+1.5,0.4);
\draw [ultra thick, orange, snake it] (0.5,0.5) -- (4.0,0.5);
\draw[ultra thick, fill=white] (0,0) rectangle (1.,1.);
\draw[ultra thick, fill=white] (\ll,0) rectangle (\ll+1,1.);
\end{tikzpicture}}\;.
\end{split}
\end{equation}
Note that Eqs.~\ref{eq:haark} and ~\ref{eq:gauss_average} are formally the same, modulo replacing the associated matrix $W$. 
In fact, in the limit of large $q$, the unitary Weingarten matrix is dominated by the diagonal contribution $\Wg_{\sigma,\pi}(q)\stackrel{q \to \infty}{=} q^{-m} \delta_{\sigma,\pi}$, implying Gauss and Haar ensemble gives consistent result upon setting the variance $\varsigma^2=q^{-1}$.

\subsection{Computations of the Transfer Matrices}
As explained in the Main Text, the key object is the generalized frame potential Eq.~\eqref{eq:frame_pot_1}. 
Using Haar invariance, we can simplify one summation setting $\pmb{z}_B'\mapsto\pmb{0}_B$, and obtain, setting $\pmb{z}\equiv\pmb{z}_B$ for notational simplicity,
\begin{equation}
        \mathcal{F}^{(k,n)} = D_B \sum_{\pmb{z}}  \mathbb{E}_{\psi} [p({\pmb{z}})^n p({\pmb{0}})^n |\braket{\tilde{\psi}(\pmb{z})|\tilde{\psi}(\pmb{0})}|^{2k}] \, .
\end{equation}

Using the MPS form Eq.~\eqref{eq:mps} of the state, we can rearrange the terms to identify the two following transfer matrices of size $\chi^2 \times \chi^2$ 
\begin{align}
    \label{eq:tmA}
\vcenter{\hbox{\includegraphics[width=0.85\linewidth]{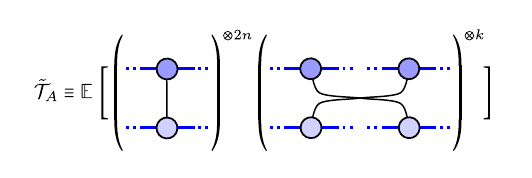}}},\\ 
\label{eq:tmB}
\vcenter{\hbox{\includegraphics[width=0.85\linewidth]{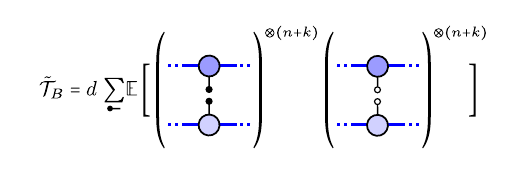} }},
\end{align}
respectively in the regions $A$ and $B$ (darker shapes represent complex conjugate tensors). Note the appearance of the permutation Eq.~\eqref{eq:miozio} in the contraction over physical indices in Eq.~\eqref{eq:tmA}. In Eqs.~\eqref{eq:tmA} and~\eqref{eq:tmB}, we used the symbol $\begin{tikzpicture}[baseline=(current  bounding  box.center), scale=0.6]
\pgfmathsetmacro{\r}{0.15}
\pgfmathsetmacro{\ll}{0.45}
\draw[line width=0.3 mm, black] (0., 0.) -- (0., \ll);
\draw[line width=0.2 mm, fill=black] (0.,0.) circle (\r); \end{tikzpicture}$ to denote a generic element $z_i$ of the local computational basis, while $\begin{tikzpicture}[baseline=(current  bounding  box.center), scale=0.6]
\pgfmathsetmacro{\r}{0.15}
\pgfmathsetmacro{\ll}{0.45}
\draw[line width=0.3 mm, black] (0., 0.) -- (0., \ll);
\draw[line width=0.2 mm, fill=white] (0.,0.) circle (\r); \end{tikzpicture}$ is the state $\ket{0}$. Finally, $d_B$ represents the local qudit dimension in region $B$, which differs between the two setups. 
The average frame potential can now be expressed as a product of transfer matrices arranged according to the circuit architecture. Specifically, for the staircase circuit (setup I), we have $\mathcal{F}^{(k,n)} = \tilde{\mathcal{T}}_A^{N_A}\tilde{\mathcal{T}}_B^{N_B}$ for the staircase circuit (setup I), while for the glued shallow circuit (setup II), the ordering alternates as $\mathcal{F}^{(k,n)} = 
    \tilde{\mathcal{T}}_B
    \tilde{\mathcal{T}}_A
    \tilde{\mathcal{T}}_B
    ... 
    \tilde{\mathcal{T}}_A \tilde{\mathcal{T}}_B$. 
We now proceed to perform the average over the random MPS tensors. Since the methodology is structurally similar in both setups, we detail the computation for setup I and defer the corresponding analysis for setup II.

Recall each tensor is a sub-block of a matrix (gate) $U$ sampled from the ensemble $\mathcal{E} \in \{ \Haar, \Gauss \}$
\begin{equation}
\vcenter{\hbox{\includegraphics[width=0.6\linewidth]{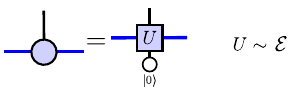}}}\;.
\end{equation}
By applying Eq.~\ref{eq:haark} and contracting the permutation operators acting on the $\chi$-dimensional auxiliary space from two adjacent sites, we derive the following transfer matrices in the replica space
\begin{equation}\label{eq:tmA_replica}
\mathcal{T}_A = 
\begin{tikzpicture}[baseline=(current bounding box.center), scale=0.75]
\tikzset{snake it/.style={decorate, decoration=snake}}
\pgfmathsetmacro{\length}{2.0} % ↓ shortened
\pgfmathsetmacro{\ll}{0.25}
\pgfmathsetmacro{\llh}{0.45}
\pgfmathsetmacro{\wr}{0.4}
\pgfmathsetmacro{\tl}{0.20}
\pgfmathsetmacro{\wl}{0.5}
\pgfmathsetmacro{\xcopy}{\length/3}
\pgfmathsetmacro{\lv}{0.8}     % ↓ shortened
\pgfmathsetmacro{\r}{0.05}
\draw [ultra thick, orange, snake it] (-0.5*\length, 0.) -- (1.2*\length, 0.); % shorter horizontal line
\draw [ultra thick, orange, snake it] (\xcopy, 0.) -- (\xcopy, \lv);         % shorter vertical line
\draw[ultra thick, fill=orange, draw=white] (\xcopy,0) circle (0.13);
\draw[line width=\wr mm, fill=white] (\xcopy-\ll,\lv-\ll) rectangle (\xcopy+\ll,\lv+\ll);
\draw[line width=\wl mm, black] (\xcopy - 1/3*\ll, \lv+\ll) -- (\xcopy - 1/3*\ll, \lv+\ll+\tl);
\draw[line width=\wl mm, black] (\xcopy+ 1/3*\ll, \lv+\ll) -- (\xcopy + 1/3*\ll, \lv+\ll+\tl);
\draw[line width=\wr mm, fill=white] (\xcopy - \ll, \lv+\ll + \tl) rectangle (\xcopy + \ll, \lv+3*\ll + \tl);
\node[scale=0.7] at (\xcopy, \lv+ 2*\ll + \tl) {$\muA$};
\node[draw, diamond, fill=orange, thick] (W) at (-\xcopy+0.4, 0.0) {};
\node[draw, rectangle, fill=blue, thick] (G) at (2.5*\xcopy, 0.0) {};
\end{tikzpicture} \;
\mathcal{T}_B = d \, \, 
\sum_{\raisebox{-0.5ex}{\raisebox{2.1ex}{\begin{tikzpicture}[scale=0.45]
    \draw[line width=0.2 mm, fill=black] (0,0) circle (0.1);
\end{tikzpicture}}}}
\begin{tikzpicture}[baseline=(current bounding box.center), scale=0.85]
\tikzset{snake it/.style={decorate, decoration=snake}}
\pgfmathsetmacro{\length}{2.0} % ↓ shortened
\pgfmathsetmacro{\ll}{0.25}
\pgfmathsetmacro{\llh}{0.45}
\pgfmathsetmacro{\wr}{0.4}
\pgfmathsetmacro{\tl}{0.20}
\pgfmathsetmacro{\wl}{0.5}
\pgfmathsetmacro{\xcopy}{\length/3}
\pgfmathsetmacro{\lv}{0.8}     % ↓ shortened
\pgfmathsetmacro{\r}{0.05}
\draw [ultra thick, orange, snake it] (-0.5*\length, 0.) -- (1.2*\length, 0.); % shorter horizontal
\draw [ultra thick, orange, snake it] (\xcopy, 0.) -- (\xcopy, \lv);         % shorter vertical
\draw[ultra thick, fill=orange, draw=white] (\xcopy,0) circle (0.13);
\draw[line width=\wr mm, fill=white] (\xcopy-\llh,\lv-\ll) rectangle (\xcopy+\llh,\lv+\ll);
\draw[line width=0.3 mm, black] (\xcopy - 2/3*\llh, \lv+\ll) -- (\xcopy - 2/3*\llh, \lv+\ll+\tl);
\draw[line width=0.3 mm, black] (\xcopy - 1/3*\llh, \lv+\ll) -- (\xcopy - 1/3*\llh, \lv+\ll+\tl);
\draw[line width=0.3 mm, black] (\xcopy+ 1/3*\llh, \lv+\ll) -- (\xcopy + 1/3*\llh, \lv+\ll+\tl);
\draw[line width=0.3 mm, black] (\xcopy+ 2/3*\llh, \lv+\ll) -- (\xcopy + 2/3*\llh, \lv+\ll+\tl);
\draw[line width=0.2 mm, fill=black] (\xcopy - 2/3*\llh, \lv+\ll+\tl) circle (\r);
\draw[line width=0.2 mm, fill=black] (\xcopy - 1/3*\llh, \lv+\ll+\tl) circle (\r);
\draw[line width=0.2 mm, fill=white] (\xcopy + 1/3*\llh, \lv+\ll+\tl) circle (\r);
\draw[line width=0.2 mm, fill=white] (\xcopy + 2/3*\llh, \lv+\ll+\tl) circle (\r);
\node[draw, diamond, fill=orange, thick] (W) at (-\xcopy+0.5, 0.0) {};
\node[draw, rectangle, fill=blue, thick] (G) at (2.5*\xcopy, 0.0) {};
\end{tikzpicture} \, ,
\end{equation}
where $\muA$ is the permutation shown in Eq.~\eqref{eq:miozio}. The orange dot in Eq.~\eqref{eq:tmA_replica} denotes the copy tensor, forcing the three incoming permutations to be the same. 
Let us now consider the two terms containing the copy tensor in Eq.~\eqref{eq:tmA_replica}.  %and \eqref{eq:tmB_replica}. 
In $A$, we identify the following diagonal matrix
\begin{equation}\label{eq:diagonal_A}
(\mathcal{D}_A)_{\pi \sigma}  =
\raisebox{2.1ex}{
\begin{tikzpicture}[baseline=(current bounding box.center), scale=0.85]
\tikzset{snake it/.style={decorate, decoration=snake}}
\pgfmathsetmacro{\length}{1.} % Length of the orange snake 
\pgfmathsetmacro{\ll}{0.25}   % Size of the T rectangles
\pgfmathsetmacro{\wr}{0.4}    % Line width T rectangles
\pgfmathsetmacro{\tl}{0.35}    % Length T lines
\pgfmathsetmacro{\wl}{0.5}    % Line width T lines
\pgfmathsetmacro{\xcopy}{0} % Length of the orange snake
\pgfmathsetmacro{\lv}{1.}     % y coordinate T rectangle
\node[scale=0.8] at (-\length - 0.2, 0) {$\pi$};
\node[scale=0.8] at (+\length + 0.2, 0) {$\sigma$};
\draw [ultra thick, orange, snake it] (-\length, 0.) -- (\length, 0.);
\draw [ultra thick, orange, snake it] (0, 0.) -- (0, \lv);
\draw[line width=\wr mm, fill=white] (\xcopy-\ll,\lv-\ll) rectangle (\xcopy+\ll,\lv+\ll);
\draw[line width=\wl mm, black] (\xcopy - 1/3*\ll, \lv+\ll) -- (\xcopy - 1/3*\ll, \lv+\ll+\tl);
\draw[line width=\wl mm, black] (\xcopy+ 1/3*\ll, \lv+\ll) -- (\xcopy + 1/3*\ll, \lv+\ll+\tl);
\draw[line width=\wr mm, fill=white] (\xcopy - \ll, \lv+\ll + \tl) rectangle (\xcopy + \ll, \lv+3*\ll + \tl);
\draw[ultra thick, fill=orange, draw=white] (0,0) circle (0.13);
\node[scale=0.7] at (\xcopy, \lv+ 2*\ll + \tl) {$\muA$};
\end{tikzpicture} } = \delta_{\pi \sigma} \, \upsilon^{(A)}_{\sigma}(d) \, ,
\end{equation}
weighting the distance of the permutation $\sigma$ from $\muA$. %Specifically, the onsite potential is $\upsilon^{(A)}_{\sigma}(d) = \llangle \sigma | \muA \rrangle_d = d^{m-\dist(\sigma,\muA)}$. 
In region $B$, instead we identify
\begin{align}\label{eq:diagonal_B}
\begin{split}
(\mathcal{D}_B)_{\pi \sigma}  &= d \, \, 
\sum_{\raisebox{-0.5ex}{\raisebox{2.1ex}{\begin{tikzpicture}[scale=0.5]
    \draw[line width=0.2 mm, fill=black] (0,0) circle (0.1);
\end{tikzpicture}}}}
\begin{tikzpicture}[baseline=(current bounding box.center), scale=0.85]
\tikzset{snake it/.style={decorate, decoration=snake}}
\pgfmathsetmacro{\length}{1.} % Length of the orange snake 
\pgfmathsetmacro{\ll}{0.25}   % Size of the T rectangles
\pgfmathsetmacro{\llh}{0.45}
\pgfmathsetmacro{\wr}{0.4}    % Line width T rectangles
\pgfmathsetmacro{\tl}{0.35}    % Length T lines
\pgfmathsetmacro{\wl}{0.5}    % Line width T lines
\pgfmathsetmacro{\xcopy}{0}    % x coordinate copy tensor
\pgfmathsetmacro{\lv}{1.}     % y coordinate T rectangle
\pgfmathsetmacro{\r}{0.05}
\node[scale=0.8] at (-\length - 0.2, 0) {$\pi$};
\node[scale=0.8] at (+\length + 0.2, 0) {$\sigma$};
\draw [ultra thick, orange, snake it] (-\length, 0.) -- (\length, 0.);
\draw [ultra thick, orange, snake it] (\xcopy, 0.) -- (\xcopy, \lv);
\draw[ultra thick, fill=orange, draw=white] (\xcopy,0) circle (0.13);
\draw[line width=\wr mm, fill=white] (\xcopy-\llh,\lv-\ll) rectangle (\xcopy+\llh,\lv+\ll);
\draw[line width=0.3 mm, black] (\xcopy - 2/3*\llh, \lv+\ll) -- (\xcopy - 2/3*\llh, \lv+\ll+\tl);
\draw[line width=0.3 mm, black] (\xcopy - 1/3*\llh, \lv+\ll) -- (\xcopy - 1/3*\llh, \lv+\ll+\tl);
\draw[line width=0.3 mm, black] (\xcopy+ 1/3*\llh, \lv+\ll) -- (\xcopy + 1/3*\llh, \lv+\ll+\tl);
\draw[line width=0.3 mm, black] (\xcopy+ 2/3*\llh, \lv+\ll) -- (\xcopy + 2/3*\llh, \lv+\ll+\tl);
\draw[line width=0.2 mm, fill=black] (\xcopy - 2/3*\llh, \lv+\ll+\tl) circle (\r);
\draw[line width=0.2 mm, fill=black] (\xcopy - 1/3*\llh, \lv+\ll+\tl) circle (\r);
\draw[line width=0.2 mm, fill=white] (\xcopy + 1/3*\llh, \lv+\ll+\tl) circle (\r);
\draw[line width=0.2 mm, fill=white] (\xcopy + 2/3*\llh, \lv+\ll+\tl) circle (\r);
\end{tikzpicture}= \delta_{\pi \sigma} \, \upsilon^{(B)}_{\sigma}(d), % \, 
\end{split}
\end{align}
which is also diagonal and favors permutations that act dis-jointly in the first and last $n+k$ replicas.
The functions $\upsilon^{(A)}_{\sigma}(d)$ and $\upsilon^{(B)}_{\sigma}(d)$ are defined in the Main Text, cf. Eq.~\eqref{eq:transfdef}. 
Finally we can rewrite $\mathcal{F}^{(k,n)}$ as the following contraction in the space of replicas
\begin{equation}\label{eq:contraction_F_app}
\mathcal{F}^{(k,n)} =  (v_L | \mathcal{T}_A^{N_A}  \mathcal{T}_B^{N_B  - 1} |v_R) \, ,
\end{equation}
where $|v_L)$, $|v_R)$ are two $m!$-dimensional boundary vectors, respectively given by 
\begin{equation}\label{eq:vr}
(v_L)_\pi=1\;,\qquad (v_R)_{\pi} = 
\begin{tikzpicture}[baseline=(current bounding box.center), scale=0.75]
\tikzset{snake it/.style={decorate, decoration=snake}} % Define the snake it style
\pgfmathsetmacro{\length}{1.} % Length of the orange snake 
\pgfmathsetmacro{\ll}{0.25}   % Size of the T rectangles
\pgfmathsetmacro{\llh}{0.45}
\pgfmathsetmacro{\wr}{0.4}    % Line width T rectangles
\pgfmathsetmacro{\tl}{0.35}   % Length T lines
\pgfmathsetmacro{\wl}{0.5}    % Line width T lines
\pgfmathsetmacro{\xcopy}{0}   % x coordinate copy tensor
\pgfmathsetmacro{\lv}{1.}     % y coordinate T rectangle
\pgfmathsetmacro{\r}{0.06}
\pgfmathsetmacro{\nl}{0.025}
\node[scale=0.8] at (-2*\length - 0.2, 0) {\large $\pi$};
\draw [ultra thick, orange, snake it] (-2*\length, 0.) -- (\xcopy, 0.);
\node[draw, diamond, fill=orange, thick] (W) at (-\length, 0.0) {};
\draw [ultra thick, orange, snake it] (\xcopy, 0.) -- (\xcopy, \lv);
\draw[ultra thick, fill=orange, draw=white] (\xcopy,0) circle (0.13);
\draw[line width=\wr mm, fill=white] (\xcopy-\llh,\lv-\ll) rectangle (\xcopy+\llh,\lv+\ll);
\draw[line width=0.3 mm, black] (\xcopy - 2/3*\llh - \nl, \lv+\ll) -- (\xcopy - 2/3*\llh - \nl, \lv+\ll+\tl);
\draw[line width=0.3 mm, blue] (\xcopy - 2/3*\llh + \nl, \lv+\ll) -- (\xcopy - 2/3*\llh + \nl, \lv+\ll+\tl);
\draw[line width=0.3 mm, black] (\xcopy - 1/3*\llh - \nl, \lv+\ll) -- (\xcopy - 1/3*\llh - \nl, \lv+\ll+\tl);
\draw[line width=0.3 mm, blue] (\xcopy - 1/3*\llh + \nl, \lv+\ll) -- (\xcopy - 1/3*\llh + \nl, \lv+\ll+\tl);
\draw[line width=0.3 mm, black] (\xcopy + 1/3*\llh - \nl, \lv+\ll) -- (\xcopy + 1/3*\llh - \nl, \lv+\ll+\tl);
\draw[line width=0.3 mm, blue] (\xcopy + 1/3*\llh + \nl, \lv+\ll) -- (\xcopy + 1/3*\llh + \nl, \lv+\ll+\tl);
\draw[line width=0.3 mm, black] (\xcopy + 2/3*\llh - \nl, \lv+\ll) -- (\xcopy + 2/3*\llh - \nl, \lv+\ll+\tl);
\draw[line width=0.3 mm, blue] (\xcopy + 2/3*\llh + \nl, \lv+\ll) -- (\xcopy + 2/3*\llh + \nl, \lv+\ll+\tl);
\draw[line width=0.2 mm, fill=black] (\xcopy - 2/3*\llh, \lv+\ll+\tl) circle (\r);
\draw[line width=0.2 mm, fill=black] (\xcopy - 1/3*\llh, \lv+\ll+\tl) circle (\r);
\draw[line width=0.2 mm, fill=white] (\xcopy + 1/3*\llh, \lv+\ll+\tl) circle (\r);
\draw[line width=0.2 mm, fill=white] (\xcopy + 2/3*\llh, \lv+\ll+\tl) circle (\r);
\end{tikzpicture}
\end{equation}

\subsection{Confinement in the staircase circuit}
We now turn to the Gaussian RMPS ensemble. 
We are interested in the scaling limit of $\mathcal{F}^{(k,n)}$ when $N_A,\chi\gg 1$ and $N_A\ll N_B \ll d^{N_A}$, while the ratio 
\begin{equation}\label{eq:scalinglimit1}
   x = \frac{d^{N_A}}{\chi}= \frac{D_A}{\chi}  \quad \mbox{(Gaussian RMPS)}, 
\end{equation}
is kept fixed.
As we will show, the results for unitary RMPS' coincide with those of the Gaussian case, provided the scaling parameter $x$ is redefined according to Eq.~\eqref{eq:scaling_1}.

\subsubsection{Leading order ($x=0$)}
At the leading order for $\chi\to\infty$, the dominat contribution is given by the diagonal $G_{\sigma \pi} (\chi) \simeq \delta_{\sigma \pi} \chi^{m}$. 
In this limit, the transfer matrices $\mathcal{T}_A$ and $\mathcal{T}_B$ are both diagonals, while the right boundary vector for the Gaussian ensemble is an equal weights superposition of all factorized permutations
\begin{equation}\label{eq:right_state_2}
    (v_R)_{\pi} \simeq \pmb{1}_F(\pi) d^2 \chi^2  \varsigma^{2m} \, .
\end{equation}
Since in $A$ the permutations are weighted by their distance from the permutation $\muA$, we now have to identify the subset of the factorized permutations $\sigma=(\sigma_1,\sigma_2)$, $\sigma_1,\sigma_2 \in S_{n+k}$,  minimizing $\dist(\sigma, \muA)$. As clarified in Ref.~\cite{Chan_2024} (see Eq. 40 therein), these permutations always act as the identity in the $n$ auxiliary replicas $\sigma_{1} = (\Id_n, \alpha)$ and $\sigma_2 = (\alpha', \Id_n)$, 
where $\alpha, \alpha' \in S_k$. Moreover the two permutations $\alpha$ and $\alpha'$ are related by an inversion, $(\Id_k, \alpha') = i_{2k} (\alpha^{-1},\Id_k) i_{2k}$. Thus, there are precisely $k!$ factorized permutations minimizing the distance from the permutation $\muA$. It can also be proven that for all of them $\dist(\sigma, \muA) = k$. 
Collecting these remarks, we have at leading order in $\chi$
\begin{align}\label{eq:zioboninoooo}
  \begin{split}
    \mathcal{F}^{(k,n,0)} &:= \lim_{\chi \to \infty} \mathcal{F}^{(k,n)}= k! \frac{\chi^2 d^{(m-k) N_A+2 N_B}}{\chi^{m(1-N)}}\varsigma^{2m N} 
  \end{split}
\end{align}
Substituting $\varsigma^2=(d\chi)^{-1}$ into Eq.~\eqref{eq:zioboninoooo} yields the frame potential for the Haar ensemble of RMPS. At leading order, this reproduces the Haar frame potential 
$\mathcal{F}^{(k,1-k,0)} = k! \, D_A^{-k} =  \mathcal{F}^{(k)}_{\Haar}$. 

\subsubsection{Domain wall picture in the scaling limit (finite $x$)}

We now consider the full calculation of $\mathcal{F}^{(k)}$, in the scaling limit Eq.~\eqref{eq:scalinglimit1}, for a fixed finite value of $x$. The discussion here for the Gaussian case is a detailed version of the one in Sec.~\ref{sec:confiI} for the unitary Haar case. Let us expand the Gram matrix in powers of $1/\chi$
\begin{equation}\label{eq:gram_matrix_exp}
    G_{\sigma \pi} (\chi) = \chi^{m} \left( \delta_{\sigma \pi}  + \sum_{\alpha\geq 1} \chi^{-\alpha} \ad_{\sigma \pi}^{(\alpha)} \right), \quad \ad_{\sigma \pi}^{(\alpha)}=\delta_{\dist(\sigma,\pi),\alpha} \;.
\end{equation}
which represents the Gaussian analogue of Eq.~\eqref{eq:Texp}. 
Note that $\ad_{\sigma \pi}^{(1)} = \ad_{\sigma \pi}$ is the adjacency matrix of permutations
introduced in the main text, indicating whether $\sigma$ and $\pi$ differ by a single transposition.  
As discussed in the main text, expanding Eq.~\eqref{eq:contraction_F_app} at first order in $1/\chi$
amounts to replace one identity with $\ad^{(1)}$ (compare with Eq.~\eqref{eq:F1basic}). This corresponds to configurations of permutations $\sigma_1, \sigma_0$ with $\dist(\sigma_1, \sigma_0) = 1$, thereby generating a single domain wall. Such a domain wall is associated with a cost $\chi^{-1}$.
Consistently with Eq.~\eqref{eq:right_state_2}, the right boundary forces $\sigma_0$ to be a factorized permutation.

To survive in the scaling limit~\eqref{eq:scalinglimit1}, the $1/\chi$ cost of the domain wall must be compensated by a sufficient number of factors $d$ obtained in the region $A$. This can happen if
$\dist(\sigma_0, \muA) = k$ and
$\dist(\sigma_1, \muA) = k-1$. 
Writing $\sigma_0 = \tau_1\ldots \tau_k \muA$, where $\{\tau_i\}$  is a set of commuting transpositions, we see that there are $k$ choices for $\sigma_1$, removing any of the $\tau_i$. At first order in $1/\chi$, this produces the same result as in \eqref{eq:F1scalinglim}. 

Let us discuss higher order in $O(\chi^{-\alpha})$ and thus of $O(x^\alpha)$ in the scaling limit. One can have insertions of $\alpha$ simple domain walls associated with $\ad^{(1)}$
at distinct bonds $\ell_{\alpha}< \ell_{\alpha-1}< \ldots< \ell_1$, each associated with the removal of a transposition $\tau \in \{\tau_i\}$ defining $\sigma_0$. There are $k!/(k-\alpha)!$ ways to choose this ordered subset of transpositions out of the $k$ transpositions $\{\tau_i\}$. However, it is also possible to have insertions of composed domain walls associated with the higher order matrices $\ad^{(p)}$. To see their effect explicitly, let us analyze in detail the case $\alpha = 2$. When the two domain walls (transpositions) occur at distinct positions $\ell_1 < \ell_2$, there are $\frac{k!}{(k-\alpha)!} = k(k-1)$ possible ordered arrangements for the transpositions $\tau_1, \dots, \tau_\alpha$. In fact, the order matters, as for instance placing $\tau_1$ at $\ell_1$ and $\tau_2$ at $\ell_2$ results in a different configuration from placing $\tau_2$ at $\ell_1$ and $\tau_1$ at $\ell_2$. In contrast, the insertion of a single matrix $\ad^{(2)}_{\tau_1 \tau_2 \muA, \muA}$ can be identified with the coinciding positions $\ell_1 = \ell_2$. Overall, we obtain
\begin{equation}
    k(k-1) \sum_{\ell_1 < \ell_2} (\dots) + \frac{k(k-1)}{2} \sum_{\ell_1 = \ell_2} (\dots) = \binom{k}{2} \sum_{\ell_1, \ell_2} (\dots) \,,
\end{equation}
where, on the right-hand side of the equality, we have extended the summation to all values of $\ell_1, \ell_2$ without constraints, thereby factorizing the combinatorial factor $\binom{k}{\alpha}$ with $\alpha = 2$. It is easy to verify from the simple form of the matrices $\ad^{(\alpha)}$ \eqref{eq:gram_matrix_exp} that this discussion generalises at higher orders, so that accounting for composite domain walls is done easily by extending the sum over $\ell_1,\ldots,\ell_\alpha$ to be unconstrained, thus including coinciding points (see also the analogous discussion in Sec. IID.2 in ~\cite{deluca2024universalityclassespurificationnonunitary}).
With these considerations in mind, we arrive at the following expansion
\begin{equation}\label{eq:expansion}
   \mathcal{F}^{(k)}(x) = \mathcal{F}^{(k)}_{\Haar} \sum_{\alpha=0}^{k} \binom{k}{\alpha} \, x^{\alpha}  f_{\alpha}  \, ,  
\end{equation}
where the sum runs over the index $\alpha$ labeling the number of domain walls. By construction, $f_{0} = 1$, as discussed in the previous section and in general 
\begin{equation}\label{eq:f_alpha}
    f_{\alpha} = \sum_{\ell_1,\ldots, \ell_{\alpha} = - \infty}^\infty d^{-V(\ell_1,\ldots, \ell_{\alpha})}
\end{equation}
where $V(\ell_1,\ldots, \ell_{\alpha}) = V_A(\ell_1,\ldots, \ell_{\alpha}) + V_B(\ell_1,\ldots, \ell_{\alpha})$ are the effective confining potential in Eq.~\eqref{eq:VA_e_VB}. 
To evaluate $f_\alpha$ as a function of $d$, let us split the sums over positive (in $B$) and negative (in $A$) $\ell$'s, defining the coefficients $a_\alpha$ and $b_\alpha$, given by $a_{0}=b_{0}=1$ and
\begin{equation}\label{eq:orchi}
\begin{split}
    a_{\alpha} &:= \sum_{\ell_1,\ldots, \ell_{\alpha}=0}^{-\infty} d^{-V_A(\ell_1,\ldots, \ell_{\alpha})}=%\sum_{\ell_1,\ldots, \ell_{\alpha}=0}^{+\infty} d^{-(\ell_1 + \ell_2 + ... + \ell_{\alpha})}= 
    \left(\frac{d}{d-1} \right)^{\alpha},\\
    b_{\alpha} &:= \sum_{\ell_1,\ldots, \ell_{\alpha} =1}^{+\infty} d^{-V_B(\ell_1,\ldots, \ell_{\alpha})}=\frac{d-1}{d} \sum_{j = 1}^{+\infty} d^{-j} j^{\alpha} \, .
\end{split}
\end{equation}
for $\alpha \geq 1$. In the above expressions, we substituted Eq.~\eqref{eq:VA_e_VB} and performed the explicit computation.  
Using Eq.~\eqref{eq:orchi}, we can decompose $f_\alpha$ 
into a sum over $\beta$, the number of domain walls in $B$
\begin{equation}\label{eq:dpG}
    f_{\alpha} = \sum_{\beta=0}^{\alpha} \binom{\alpha}{\beta} a_{\alpha-\beta} b_{\beta} = \left(\frac{d}{d-1}\right)^{\alpha-1} \left(1 + \sum_{j = 1}^{+\infty} d^{-j}    \big( 1 + j \frac{d-1}{d} \big)^{\alpha} \right) \, ,
\end{equation}
where $\beta$ is the number of particles in $B$. 
Plugging this expression in Eq.~\eqref{eq:expansion}, we arrive at
\begin{equation}\label{eq:mustardddd}
   \mathcal{F}^{(k)}(x) = \mathcal{F}^{(k)}_{\Haar} \frac{d-1}{d} \sum_{j = 0}^{+\infty} d^{-j} \left(1 + x \frac{d}{d-1} + x j \right)^k \, .
\end{equation}
Recall that $\mathcal{F}^{(k)}$ is, up to rescaling, the $k$-th moment of the random variable $u$, distributed according to Eq.~\eqref{eq:prob_distr}. 
To recover the distribution, it is useful to identify the moments as obtained from the decomposition $u=u_0 z$ in terms of two independent random variables: a Porter-Thomas distributed variable $u_0$ with probability $p_{u_0}(u_0)=e^{-u_0}$ and moments $\Ex[u_0^k]= D_A^{k} \mathcal{F}^{(k)}_{\Haar} = k!$, and a discrete random variable $z$, with distribution $p_z(z) =(d-1) \sum_{j = 0}^{+\infty} d^{-j-1}
    \delta(z-z_{j})$ and moments $\Ex[z^k]=(d-1)\sum_{j=0}^{+\infty}d^{-j+1} (z_j)^k$ where $z_{j} = 1 + x \frac{d}{d-1} + x j$. 
As a result, the distribution of $u$ is the multiplicative convolution of $p_{u_0}$ and $p_z$, namely
\begin{equation}
     P(u; x) = \int_{-\infty}^{+\infty} \frac{dt}{|t|} p_{u_0}(t) p_z \left(\frac{u}{t} \right)=\frac{d-1}{d} \sum_{j=0}^{\infty} e^{- \frac{u}{z_{j}}} z_{j}^{-1}  d^{-j} \, .
\end{equation}
Note that $P(u; x)$ is properly normalized, since $\int_0^{\infty} du \, e^{- \frac{u}{z_{j}}} z_{j}^{-1} = 1$ and %thus $\int_0^{\infty} du P(u; x) = 
$\frac{d-1}{d}\sum_{j=0}^{\infty} d^{-j} = 1$, and depends parametrically on the scaling variable $x$. In particular, for $x\to0$ we recover the Porter-Thomas distribution $\lim_{x \to 0^+} P(u;x) =e^{-u}$ since $\lim_{x \to 0^+} z_j=1$ for all $j$. 

\subsubsection{Unitary case}
We can now extend our results to the ensemble of Haar unitary RMPS. This extension requires to handle the non-trivial expression of the Weingarten matrix in Eqs.~\ref{eq:tmA_replica}. Crucially, the product of transfer matrices $\mathcal{T}_A$, $\mathcal{T}_B$ depends only on $T(\chi, d) = W(d\chi) G(\chi)$. 
The Weingarten matrix acts as a dressing of the ferromagnetic interaction between adjacent sites 
in the Gaussian case encoded in $G(\chi)$. Remarkably, it is still possible to obtain a closed expression in the scaling limit. To see this, we note that in the scaling limit \eqref{eq:scaling_1}, as we discussed in~\ref{sec:confiI}, the relevant configurations in the scaling limit at the order $O(x^\alpha)$ contain $\alpha$ domain walls at positions $\ell_\alpha, \ell_{\alpha-1},\ldots, \ell_1$; thus, they are made of
$\alpha+1$ domains, $\sigma_{\alpha},\sigma_{\alpha-1},\ldots,\sigma_1, \sigma_{0} = \sigma$ with $\sigma \in \Sfact$ and $\dist(\sigma_{a}, \muA) = k - a$. As discussed in the previous section, higher orders in the $\chi \to \infty$ expansion of $T(\chi, d)$ can be associated with cases where  $\ell_i = \ell_{i+1} = \ldots = \ell_{i+\beta-1}$. This corresponds to the composite domain wall between $\sigma_{i + \beta - 1}$ and $\sigma_{i-1} = \tau_{m_1}\ldots \tau_{m_\beta} \sigma_{i + \beta - 1}$, so that $\dist(\sigma_{i-1}, \sigma_{i + \beta - 1}) = \beta$. 
The associated cost is then given by $T_{\sigma_{i + \beta - 1}, \sigma_{i - 1}}(\chi, d) = \chi^{-\beta} d^{-m} \mathsf{T}^{(\beta)}_{\sigma_{i + \beta - 1}, \sigma_{i - 1}} + O(\chi^{-\beta - 1})$. This term can be explicitly computed from the known expansion of Weingarten functions. Specifically, setting $\sigma = \sigma_{i + \beta - 1}$ and $\sigma' = \sigma_{i - 1} = \tau_{m_1}\ldots \tau_{m_\beta} \sigma$, we have
\begin{equation}
    T_{\sigma, \sigma'}(\chi, d) = 
    \sum_{\pi \in S_m} [W (d \chi)]_{\sigma' \pi} [G(\chi)]_{\pi \sigma} \;.
\end{equation}
At the leading order $O(\chi^{-\beta})$, the sum over $\pi$ is dominated by the cases when the intermediate permutation $\pi = \tau_{g_1} \ldots \tau_{g_b} \sigma$, where $\{\tau_{g_i}\}_{i=1}^b$ are a subset of size $b$ of $\{\tau_{m_i}\}_{i=1}^\beta$. This implies 
\begin{equation}
    [\Wg(d \chi)]_{\sigma' \pi} = (\chi d)^{-m}\left(\frac{(-1)^{\beta - b}}{d^{\beta - b} \chi^{\beta - b}} + O((d \chi)^{\beta - b+1})\right)
\end{equation}
Finally, since $d(\sigma, \pi) = b$, we have $G_{\pi \sigma}(\chi) = \chi^{m - b}$. Summing over the possible choices of the $b$ transpositions among the $\beta$ ones,
we obtain 
\begin{equation}
    \mathsf{T}^{(\beta)}_{\sigma' \sigma} = \sum_{b=0}^\beta \binom{\beta}{b} \frac{(-1)^b}{d^b} = \big(\frac{d-1}{d}\big)^\beta
    \end{equation}
which shows that $(d-1)/d$ effectively renormalizes the interaction strength.
Overall, this amounts only to redefining the scaling variable $x\mapsto x (d-1)/d$, consistent with Eq.~\eqref{eq:scaling_1} in the Main Text.

\def\bibsection{\section*{References}} 
\bibliography{fildered_bib}

\section*{Acknowledgments}
We thank M. McGinley, W.W. Ho, L. Piroli, P. Sierant, and A. Lerose for inspiring discussions. 
J.D.N. and G.L. are founded by the ERC Starting Grant 101042293 (HEPIQ) and the ANR-22-CPJ1-0021-01. ADL acknowledges support from the
ANR JCJC grant ANR-21-CE47-0003 (TamEnt). X.T. acknowledges support from DFG under Germany's Excellence Strategy – Cluster of Excellence Matter and Light for Quantum Computing (ML4Q) EXC 2004/1 – 390534769, and DFG Collaborative Research Center (CRC) 183 Project No. 277101999 - project B01. 

\section*{Data Availability}
The data for the results presented in this work will be publicly available at publication. 

\section*{Author contributions}
G. Lami performed the numerical simulation and contributed to the analytical computations. A. De Luca contributed to the analytical computations. 
All authors equally co-designed the project and wrote the paper. 

\section*{Competing Interests}
The authors declare no competing interests.

\newpage
\newpage

\section{Supplementary Note I: Details of the computation for the glued random tensor network}
In this case, the unitary defining the random matrix product state are given by 
\begin{equation}
\begin{split}
&\vcenter{\hbox{\includegraphics[width=0.6\linewidth]{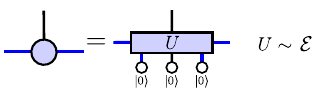}}}\,\text{ in region $A$} \\
\label{eq:tmA_2_1}
&\vcenter{\hbox{\includegraphics[width=0.6\linewidth]{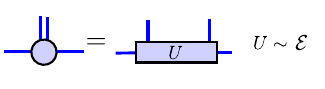}}} \, \text{ in region $B$}.
\end{split}
\end{equation}
As described in Methods, using the Weingarten calculus and contracting, we obtain the following two transfer matrices
\begin{equation}\label{eq:tmA_2_1}
\vcenter{\hbox{\includegraphics[width=0.85\linewidth]{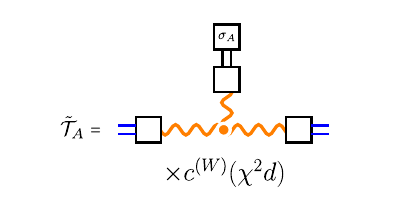}}}
\end{equation}
where the constant 
\begin{align}\label{eq:constant_cW}
\begin{split}
    c^{(W)}(q) &= \sum_{\pi} \Wg_{\sigma,\pi}{(q)} = \\ 
    &=\begin{cases}
        \varsigma^{2m}&\,  \, \text{ when } \,  \, \mathcal{E}=\Gauss\\
        \frac{1}{q (q+1) ... (q+m-1)} &\,  \, \text{  when   } \,  \, \mathcal{E}=\Haar \, ,
    \end{cases}    
\end{split} 
\end{align}
comes from contracting with the bottom replica $|0\rangle$ states. 

Similarly, the transfer matrix in region $B$ reads
\begin{equation}
\vcenter{\hbox{\includegraphics[width=0.9\linewidth]{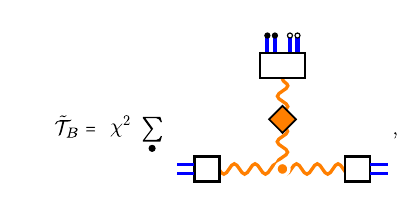}}} \, 
\end{equation}
where we used that the local physical dimension (which is affected by measurements) is $d_B=\chi^2$. 

Concatenating the matrices $\tilde{\mathcal{T}}_A$, $\tilde{\mathcal{T}}_B$ yields, as before, new transfer matrices acting in the space of replicas
\begin{equation}\label{eq:tmA_2_replicas}
\vcenter{\hbox{\includegraphics[width=0.8\linewidth]{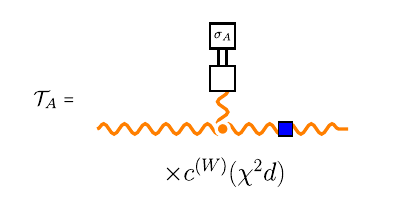}}}
\end{equation}
\begin{equation}\label{eq:tmB_2_replicas}
\vcenter{\hbox{\includegraphics[width=0.8\linewidth]{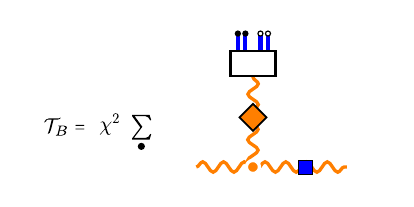}}}
\end{equation}

We note that the diagonal matrix in region $A$ is the same as in Setup $I$, cf. Methods. 
Instead $\mathcal{D}_B$ for Setup $II$ is given by
\begin{align}\label{eq:diagonal_B_new}
\begin{split}
&(\mathcal{D}_B)_{\pi \sigma}  = \chi^2 \, \, 
\sum_{\raisebox{-0.5ex}{\raisebox{2.1ex}{\begin{tikzpicture}[scale=0.5]
    \draw[line width=0.2 mm, fill=black] (0,0) circle (0.1);
\end{tikzpicture}}}}
\begin{tikzpicture}[baseline=(current bounding box.center), scale=0.85]
\tikzset{snake it/.style={decorate, decoration=snake}}
\pgfmathsetmacro{\length}{1.} % Length of the orange snake 
\pgfmathsetmacro{\ll}{0.25}   % Size of the T rectangles
\pgfmathsetmacro{\llh}{0.45}
\pgfmathsetmacro{\wr}{0.4}    % Line width T rectangles
\pgfmathsetmacro{\tl}{0.35}    % Length T lines
\pgfmathsetmacro{\wl}{0.5}    % Line width T lines
\pgfmathsetmacro{\xcopy}{0}    % x coordinate copy tensor
\pgfmathsetmacro{\lv}{2.}     % y coordinate T rectangle
\pgfmathsetmacro{\r}{0.05}
\node[scale=0.8] at (-\length - 0.2, 0) {$\pi$};
\node[scale=0.8] at (+\length + 0.2, 0) {$\sigma$};
\node[scale=0.8] at (0.4, \lv/2+0.4) {$\sigma'$};
\draw [ultra thick, orange, snake it] (-\length, 0.) -- (\length, 0.);
\draw [ultra thick, orange, snake it] (\xcopy, 0.) -- (\xcopy, \lv);
\draw[ultra thick, fill=orange, draw=white] (\xcopy,0) circle (0.13);
\node[draw, diamond, fill=orange, thick] (W) at (\xcopy, \lv/2) {};
\draw[line width=\wr mm, fill=white] (\xcopy-\llh,\lv-\ll) rectangle (\xcopy+\llh,\lv+\ll);
\draw[line width=0.3 mm, ultra thick, blue] (\xcopy - 2/3*\llh, \lv+\ll) -- (\xcopy - 2/3*\llh, \lv+\ll+\tl);
\draw[line width=0.3 mm, ultra thick, blue] (\xcopy - 1/3*\llh, \lv+\ll) -- (\xcopy - 1/3*\llh, \lv+\ll+\tl);
\draw[line width=0.3 mm, ultra thick, blue] (\xcopy+ 1/3*\llh, \lv+\ll) -- (\xcopy + 1/3*\llh, \lv+\ll+\tl);
\draw[line width=0.3 mm, ultra thick, blue] (\xcopy+ 2/3*\llh, \lv+\ll) -- (\xcopy + 2/3*\llh, \lv+\ll+\tl);
\draw[line width=0.2 mm, fill=black] (\xcopy - 2/3*\llh, \lv+\ll+\tl) circle (\r);
\draw[line width=0.2 mm, fill=black] (\xcopy - 1/3*\llh, \lv+\ll+\tl) circle (\r);
\draw[line width=0.2 mm, fill=white] (\xcopy + 1/3*\llh, \lv+\ll+\tl) circle (\r);
\draw[line width=0.2 mm, fill=white] (\xcopy + 2/3*\llh, \lv+\ll+\tl) circle (\r);
\end{tikzpicture}=\\
&= \delta_{\pi \sigma} \, \sum_{\sigma'} \Wg_{\sigma',\sigma}{(\chi^2)}  \bigg(\pmb{1}_F(\sigma') \chi^4 + \big(1 -\pmb{1}_F(\sigma') \big) \chi^2 \bigg) \, .
\end{split}
\end{align}
Note that Eq.~\eqref{eq:diagonal_B_new} contains a term that favors factorized permutations, assigning them a weight of $\chi^4$ compared to the cost $\chi^2$ for non-factorized permutations. This term, which acts analogously to an external magnetic field, now appears `dressed' by the Weingarten coefficients. 

As before, we can write $\mathcal{F}^{(k,n)}$ as product of transfer matrices in the replica space. Specifically 
\begin{equation}\label{eq:contraction_F_new}
\mathcal{F}^{(k,n)} = (v_L| (\mathcal{T}_A \mathcal{T}_B)^{N_A-1}  \mathcal{T}_A |v_R) \, ,
\end{equation}
where $|v_L)$, $|v_R)$ are $m!$-boundary vectors given by %Specifically, with the circuit geometry shown in Eq.~\ref{eq:rmps2}, one gets 
\begin{equation}\label{eq:vLvR_2_replicas}
\vcenter{\hbox{\includegraphics[width=0.85\linewidth]{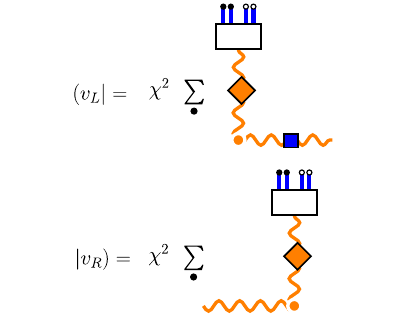}}} \,. 
\end{equation}

\section{Confinement in the glued RMPS (setup $II$)}
We now present the detailed computation of the generalized frame potential $\mathcal{F}^{(k,n)}$. As outlined in the Methods section, we begin with the Gaussian ensemble, which allows for analytical tractability, and later recover the corresponding result for the Haar unitary ensemble.

The relevant scaling limit for the Gaussian case is given by
\begin{equation}\label{eq:scalinglimit2}
x = \frac{N_A}{\chi^2} \;.
\end{equation}
Up to a rescaling of the parameter $x$, the same scaling behavior governs the unitary case as well, indicating that the qualitative features of the frame potential are preserved across both ensembles.

\subsubsection{Leading order ($x=0$)}
At the leading order for $\chi \rightarrow \infty$, the terms coming from factorized permutations which minimize the distance from the permutation $\muA$ are dominants. Considering these vacuum states, we get  
\begin{align}
  \begin{split}
    \mathcal{F}^{(k,n,0)} &= k! (\varsigma_A^{2m} d^{m - k})^{N_A} (\varsigma_B^{2m} \chi^{4})^{N_A+1} (\chi^m)^{2 N_A} \, ,
  \end{split}
\end{align}
and, setting the Gaussian variance to $\varsigma^{2}_A = (d \chi^2)^{-1}$ for the gates acting on sites $A$ and $\varsigma^{2}_B = (\chi^2)^{-1}$ for the sites $B$, gives
\begin{align}
  \begin{split}
    \mathcal{F}^{(k,n,0)} &= k! \big(\frac{1}{d^m \chi^{2m}} d^{m - k} \big)^{N_A} \big(\frac{1}{\chi^{2m}} \chi^{4} \big)^{N_A+1} (\chi^m)^{2 N_A} = \\ &= k! d^{- k N_A} \chi^{(4 - 2m)(N_A+1)} \, .     
  \end{split}
\end{align}
In the replica limit $m \rightarrow 2$, and still for $\chi \rightarrow \infty$ we recover therefore 
the Haar value
\begin{align}
  \begin{split}
    \mathcal{F}^{(k,1-k,0)} = k! \, D_A^{-k} =  \mathcal{F}^{(k)}_{\Haar}  \, .  
  \end{split}
\end{align}

\subsubsection{Domain wall picture in the scaling limit (finite $x$)}
Next, we provide a complete classification of all possible excitation types that can emerge above the $k!$ vacuum states. For clarity, we introduce distinct labels for each class of excitations.

\begin{figure}[ht]
%\vcenter{\hbox{
\includegraphics[width=1.\linewidth]{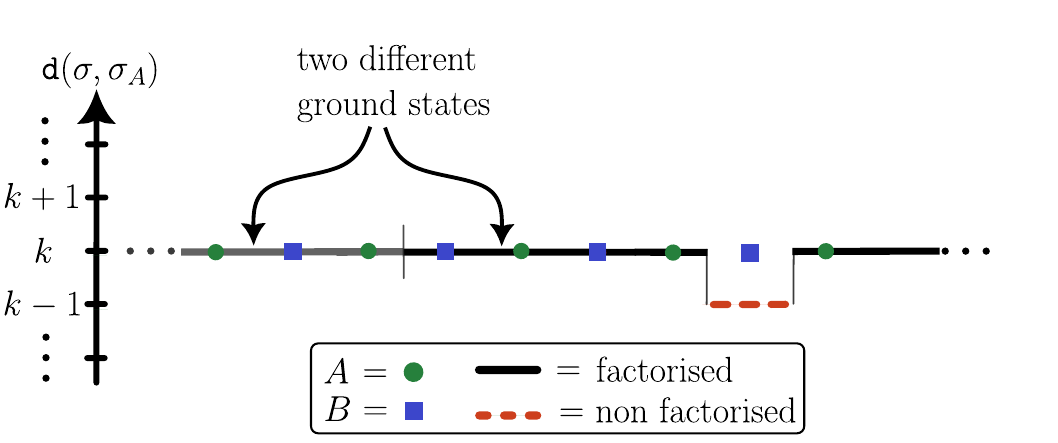}
%}}
\caption{A graphical representation of domain-walls excitations and confinement for glued RMPS (setup $II$). We show excitations of type \textbf{F2} (left) and A2 (right).
\label{fig:eccitazioni_4}
}
\end{figure}

\begin{enumerate}
\item[\textbf{A1}] These excitations are localized on a single site in region \( A \). A Gram matrix insertion can induce a transposition that reduces the distance between the current permutation \( \sigma \) and \( \mu_A \) from \( k \) to \( k - 1 \). This operation incurs a cost of \( 1/\chi \) and has multiplicity \( k \), corresponding to the choice of any one of the \( k \) transpositions that separate \( \sigma \) from \( \mu_A \). To avoid a penalty of order \( \chi^2 \) on the subsequent \( B \) site---since the resulting permutation is no longer factorized---the excitation must be removed immediately after the \( A \) site. This contributes an additional factor of \( 1/\chi \). Altogether, these excitations yield a contribution of order \( \frac{N_A}{\chi^2} \, k d \).
    \item[\textbf{F1}] These excitations transition one vacuum state, $\sigma$, to another factorized permutation, which must necessarily lie at distance  $k+1$ from $\muA$. The excitations can begin and end at any site, incurring an energy cost of $1/d$ at each involved $A$ sites. Therefore, we have to sum over all possible lengths $\ell$ of the excitation. 
    When the excitation originates before a $B$ site, the contribution is $1 + 2 \sum_{\ell=1}^{\infty} d^{-\ell}$, while an excitation starting before an $A$ site yields $2 \sum_{\ell=1}^{\infty} d^{-\ell}$. Combining these two cases, we obtain the total factor $1 + 4 \sum_{\ell=1}^{\infty} d^{-\ell} = \frac{d+3}{d-1}$, which accounts for all energetic costs on $A$ sites. 
    The permutation multiplicity of these excitations is $2 \frac{m}{2} (\frac{m}{2} - 1)$, since one can choose one of the $\frac{m}{2} (\frac{m}{2} - 1)$ transpositions that leave the permutation factorized and after it can return with either the same transposition or the inverse. We therefore get a contribution $\frac{N_A}{\chi^2} \frac{d+3}{d-1} 2 \frac{m}{2} (\frac{m}{2} - 1)$.
    \item[\textbf{F2}] These excitations are related to the contact between two distinct vacuum states $\sigma',\sigma$ at a distance $\dist(\sigma',\sigma)=2$ from each other. Locally, the Gram matrix makes the vacuum have a double jump (two transpositions) and transition to another vacuum state. The cost of the domain wall is $1/\chi^2$. The multiplicity can be counted as in the previous point, but with $m$ replaced by $k$, since all the permutations involved are factorized and act as identity on the two sets of $n$ elements. Consequently, we get $\frac{N_A}{\chi^2} \frac{2}{2} k (k-1)$.
     \item[\textbf{A2}] Finally, there is an additional type of excitations involving a single site $A$. They bring a vacuum state to a permutation (not necessarily factorized) at distance $k+1$ from $\muA$, paying a cost $1/d$. The multiplicity is the total number of transpositions $(m (m - 1))/2$ minus the transposition considered in the excitations \textbf{A1}, minus the one considered in F1 (specifically, that of length $\ell=1$). We therefore get $\frac{N_A}{\chi^2} \frac{1}{d} \left(\frac{m (m - 1)}{2} - k - \frac{m}{2} (\frac{m}{2} - 1) \right)$.
\end{enumerate}
Excitations of type \textbf{F1} and \textbf{A1} are illustrated in the Main Text, while excitations of type \textbf{F2} and \textbf{A2} are shown here in Fig.\ref{fig:eccitazioni_4}.

The first-order correction in $x$ obtained by summing all considered excitations, explicitly 
\begin{widetext}
\begin{align}
    \begin{split}
    x \bigg(  k d &+ \frac{1}{d} \big(\frac{m (m - 1)}{2} - k - \frac{m}{2} (\frac{m}{2} - 1) \big) + m (\frac{m}{2} - 1) \frac{d+3}{d-1} + k (k - 1) \bigg) \, .
    \end{split}
\end{align}
In the scaling limit under consideration, these excitations are dilute and can be created independently. As a result, to capture contributions at all orders in the \( x \)-expansion, we may exponentiate the expression above. The resulting formula for the generalized frame potential in the Gaussian ensemble is 
\begin{align}
    \begin{split}
    \mathcal{F}^{(k,n)} = \mathcal{F}^{(k,n,0)} \exp &\bigg[ x \bigg(  k d + \frac{1}{d} \big(\frac{m (m - 1)}{2} - k - \frac{m}{2} (\frac{m}{2} - 1) \big) + m (\frac{m}{2} - 1) \frac{d+3}{d-1} + k (k - 1) \bigg) \bigg]   \,.
    \end{split}
\end{align}

\subsubsection{Unitary case}
In the unitary case, other two contributions arise in the expansion of the generalized frame potential:
\begin{enumerate}
    \item [\textbf{U1}] The Weingarten constant $c^{(W)}(q)$ appearing at sites $A$ (see Eqs.~\eqref{eq:tmA_2_1} and \eqref{eq:constant_cW}) can be expanded as 
    \begin{equation}
    c^{(W)}(q) = q^{-m} \left(1-\frac{m(m-1)}{2 q}+o\left(q^{-2}\right)\right) \, ,
    \end{equation}
    for large $q$. Since this constant multiplies the transfer matrix at all sites $A$ with $q=d\chi^2$, we get an overall factor $\left( c^{(W)}(d \chi^2) \right)^{N_A}$. Expanding this factor, one gets a term $(d \chi)^{2 N_A} (1-\frac{m(m-1)}{2 d \chi^2} N_A) =(d \chi)^{2 N_A} (1-x \frac{m(m-1)}{2 d})$. We therefore get a new contribution $-x \frac{m(m-1)}{2 d}$ in our scaling limit.
    \item [\textbf{U2}] In the sites $B$, the Weingarten matrix $\Wg(\chi^2)$ that appears in Eq.\eqref{eq:diagonal_B_new} can insert a transposition 
    and let one vacuum state $\sigma$ transition to another factorized permutation $\sigma'$ at distance $1$. Specifically, the transposition corresponds to the insertion of adjacency matrix $\ad_{\sigma' \sigma}$ in the Weingarten expansion $W_{\sigma' \sigma} (\chi) = (\chi)^{-m} \left( \delta_{\sigma' \sigma}  - \ad_{\sigma' \sigma} (\chi)^{-1} + O(\chi^{-2}) \right)$.
    There are $\frac{m}{2} (\frac{m}{2} - 1 )$ of selecting the transposition. We therefore get a new contribution $-x \frac{m}{2} (\frac{m}{2} - 1 )$.  
\end{enumerate}

Overall, we thus obtain
\begin{align}
    \begin{split}
    \mathcal{F}^{(k,n)} = \mathcal{F}^{(k,n,0)} \exp &\bigg[ x \bigg(  k d + \frac{1}{d} \big(\frac{m (m - 1)}{2} - k - \frac{m}{2} (\frac{m}{2} - 1) \big) +  m (\frac{m}{2} - 1) \frac{d+3}{d-1} + k (k - 1)  -\frac{(m-1) m}{2 d}-\frac{m}{2} \left(\frac{m}{2}-1\right)\bigg) \bigg]  \,.
    \end{split}
\end{align}
Finally taking the replica limit $n \to 1-k$ ($m \to 2$), we recover the Main Text result, namely
\begin{align}
    \begin{split}
    \mathcal{F}^{(k,1-k)} = \mathcal{F}^{(k)}_{\Haar} \exp \bigg[ x \bigg(  k^2 + k \frac{d^2 - d - 1}{d} \bigg) \bigg]  \,,
    \end{split}
\end{align}
\end{widetext}

\end{document}